\documentclass[english,11pt,a4paper,oneside]{article}

\usepackage[utf8]{inputenc}
\usepackage{amssymb}
\usepackage{amsmath}
\usepackage{braket}
\usepackage{authblk}
\usepackage[top=2.5 cm, bottom=2.5 cm, left=2.5 cm, right=2.5 cm]{geometry}
\usepackage{graphicx}
\usepackage[table,xcdraw]{xcolor}
\usepackage{subcaption}
\usepackage{comment}
\usepackage[numbers,sort&compress]{natbib}
\usepackage{bm}
\usepackage{hyperref}
\usepackage{multirow}
\usepackage[normalem]{ulem}

\DeclareMathOperator{\sech}{sech}
\newcommand \lan {\langle}
\newcommand \ran {\rangle}
\newcommand{\be}{\begin{equation}}
\newcommand{\ee}{\end{equation}}
\newcommand{\bea}{\begin{eqnarray}}
\newcommand{\eea}{\end{eqnarray}}
\newcommand \e {\textrm e}

\def\p#1{\textcolor{purple}{#1}}
\newcommand{\pa}[3]{\ensuremath{\frac{\partial^{#3} {#1}}{\partial {#2}^{#3}}}}

\begin{document}

\title{Signature of (anti)cooperativity in the stochastic fluctuations of small systems: application to the bacterial flagellar motor}
	
\author[1]{María-José Franco-Oñate}
\author[1]{Andrea Parmeggiani}
\author[1]{Jérôme Dorignac}
\author[1]{Frédéric Geniet}
\author[1]{Jean-Charles Walter}
\author[2]{Francesco Pedaci}
\author[2]{Ashley L Nord}
\author[1,*]{John Palmeri}
\author[1,*]{Nils-Ole Walliser}

\affil[1]{\footnotesize{Laboratoire Charles Coulomb (L2C), Univ. Montpellier, CNRS, Montpellier, France}}
\affil[2]{Centre de Biologie Structurale (CBS), Univ. Montpellier, CNRS, INSERM, Montpellier, France}

\affil[*]{Corresponding authors: john.palmeri@umontpellier.fr, nils-ole.walliser@umontpellier.fr}

\date{}

\bibliographystyle{ieeetr}

\maketitle

\begin{abstract}
\noindent The cooperative binding of molecular agents onto a substrate is pervasive in living systems. 
To study whether a system shows cooperativity, one can rely on a fluctuation analysis of quantities such as the number of substrate-bound units and the residence time in an occupancy state.
Since the relative standard deviation from the statistical mean monotonically decreases with the number of binding sites, these techniques are only suitable for small enough systems, such as those implicated in stochastic processes inside cells.
Here, we present a general-purpose grand canonical Hamiltonian description of a small one-dimensional (1D) lattice gas with either nearest-neighbor or long-range interactions as prototypical examples of cooperativity-influenced adsorption processes. 
First, we elucidate how the strength and sign of the interaction potential between neighboring bound particles on the lattice determine the intensity of the fluctuations of the mean occupancy.
We then employ this relationship to compare the theoretical predictions of our model to data from single molecule experiments on bacterial flagellar motors (BFM) of \textit{E. coli}. 
In this way, we find evidence that cooperativity controls the mechano-sensitive dynamical assembly of the torque-generating units, the so-called stator units, onto the BFM. 
Furthermore, in an attempt to quantify fluctuations and the \textit{adaptability} of the BFM, we estimate the stator-stator interaction potential. 
Finally, we conclude that the system resides in a \textit{sweet spot} of the parameter space (\textit{phase diagram}) suitable for a smoothly adaptive system while minimizing fluctuations.
\end{abstract}

\tableofcontents
	
\section{Introduction}

Cooperativity is a pervasive phenomenon that emerges in biological systems at different time and length scales,
and levels of complexity \cite{whitty2008cooperativity}: from the molecular scale \cite{Ogata1972, Dill1993, Tielrooij2010,williamson2008cooperativity}  to multiple-cell organization \cite{West2007}, extending to the emergence of collective behavior in many-body systems. 

We focus here on cooperative processes involving the adsorption of ligands onto a substrate disposing of a limited number of binding sites \cite{owen2023size}.
Our main goal is to develop a general method for assessing how short- and long-range interactions between substrate-bound ligands affect stochastic fluctuations in the number of adsorbed units.
In the presence of cooperativity, small system size effects strongly influence the amplitude of the stochastic fluctuations and the shape of the equilibrium probability distribution function (PDF) describing substrate occupancy.
By studying these characteristic signatures of cooperativity, we propose 1) a criterion to determine whether any given adsorption system exhibits cooperative or anti-cooperative behavior and 2) a method to quantify the amplitude of the ligand-ligand interaction potential.

To this end, we introduce a minimal 1D lattice gas model with interacting units in equilibrium with a thermal bath and chemical reservoir, described using a general-purpose grand canonical Hamiltonian with short or long-range interactions. 
With it, we study how fluctuations at equilibrium, described by the standard deviation of the occupancy, are influenced by the model parameters, namely the ligand binding energy, the ligand-ligand interaction potential, and the chemical potential of the bulk reservoir.
To get closer to observable quantities, we invert the problem by determining the key formulae relating the occupancy standard deviation and PDF to the experimentally accessible average occupancy of the system. 
In this way, we are able to quantify how, at fixed average occupancy, increasing cooperativity leads to an increase in fluctuations and how increasing anti-cooperativity leads to a decrease.
In the process, we highlight the saturation of fluctuations for both strong cooperativity and strong anti-cooperativity, as well as a subtle difference between fluctuations near half-filling for systems with even and odd binding sites in the case of strong anti-cooperativity.
Finally, we apply the model by comparing our theoretical predictions with experimental data for the bacterial flagellar motor (BFM) of \textit{E. coli} \cite{Berg2003,Nirody2017}, a macro-molecular complex that has previously been found to exhibit cooperative binding of the torque-generating (stator) units \cite{Ito2021}. 
In this previous work, although evidence for cooperativity was uncovered in stator unit assembly dynamics, no estimate of the stator-stator interaction potential was proposed, which is our goal here. 
In Methods Section \ref{ssec:methods-Hillcoeff}, we make the connection between our model-based approach and phenomenological approaches based on effective Hill coefficients \cite{owen2023size,Bai2010}.

We also discuss the concept of motor adaptability to changes in external conditions and set up a general framework to assess its desired level depending on how the motor is expected to function. 
Our fluctuation analysis provides evidence for cooperativity by leading to an estimate of the stator-stator interaction potential between 1 and 2 $k_\text{B}T$ for the short-range  model, a result that is coherent with this system's expected smooth adaptive nature to external stimuli \cite{Fukuoka2009,Tipping2013,Castillo2013,Wadhwa2019,Nirody2019}.
We contrast this moderate cooperative behavior found for stator unit binding to the BFM with the highly cooperative behavior of rotor-switching units that are believed to be behind the rapid stochastic switching in BFM rotational direction \cite{Bai2010,Bai2012Coupling}.
	
\section{Results}
In this section, we propose a minimal prototypical model of cooperative particle binding on a substrate of finite size.
We then obtain an analytical expression for the fluctuations of the mean occupancy as a function of the relevant model parameters, including the interaction potential between bound particles, which acts as a proxy for system cooperativity.
Finally, we apply these results to investigate the characteristic signatures of cooperativity by establishing a general criterion for its manifestation and then propose a procedure to estimate the interaction potential from experimental data.
	
\subsection{Lattice gas model}
Consider a periodic 1D lattice with $L$ binding sites in contact with a heat and particle reservoir of constant temperature $T$ and effective chemical potential $\mu$.
We assign to each binding site $i$ ($i=1,..., L$) a binary variable $\varphi_i$: $\varphi_i=1$ means that a particle (ligand) occupies the $i$-th site; otherwise, $\varphi_i=0$. 
The array $\bm{\varphi}=\{ \varphi_i \} = \left(\varphi_1,...,\varphi_L\right)$ uniquely defines one of the $2^L$ possible \textit{microscopic} configurations (or \textit{microstates}). 
The relative occupancy in a given microstate is $\varphi = L^{-1} \sum_{i=1}^L\varphi_i$.
We say that the system is in \textit{mesostate} $N$ if the number of particles on the lattice, i.~e. the occupancy of the system equals $N = L\varphi$.
There are $C^L_N = L! /\left[N! \left(L-N\right)!\right]$ microstates $\bm{\varphi}$ consistent with the occupancy $N$; in this case, we say that the mesostate $N$ has multiplicity $C^L_N$.

The energy of the system  depends on the specific microscopic configuration $\bm{\varphi}$ according to the grand canonical effective 1D short-range interaction Hamiltonian:
	\begin{equation}\label{eq:Hamiltonian}
	\mathcal{\beta H(\bm{\varphi})} = -J  \sum_{i=1}^{L} \varphi_i \varphi_{i+1} - \mu \sum_{i=1}^{L} \varphi_{i}.
	\end{equation}
Since we choose periodic boundary conditions, \mbox{$i= L+1$} corresponds to $i=1$.
Furthermore, we introduce $\beta = 1/(k_\text{B}T)$ with the Boltzmann constant $k_\text{B}$, rendering the two adjustable control parameters of our model, $J$ and $\mu$, dimensionless. The effective chemical potential, 
$\mu$, is considered to be made up of two parts,  
$\mu = \mu_{\rm r} - \varepsilon$, where $\mu_{\rm r}$ is the chemical potential of the external particle reservoir and $\varepsilon < 0$ is the binding energy of particle adsorption onto the substrate. 
The nearest-neighbor interaction $J$ is attractive for $J>0$ (cooperativity) and repulsive for negative values (anti-cooperativity).  
For fixed $J$ the number of bound particles can be modulated by varying $\mu$  through $\mu_{\rm r}$ and/or $\varepsilon$.
When $J=0$ we recover the usual non-interacting (Hill-Langmuir) model.

The predictions for this short-range model will also be compared below with those for the long-range model where all pairs of particles interact with the same strength independent of their relative distance on the lattice.
	
\subsection{Mean occupancy and its standard deviation}
The Hamiltonian  Eq.~(\ref{eq:Hamiltonian}) falls into the universality class of the 1D short-range lattice gas model.
Hence, we can map our model to the 1D Ising model (see \cite{mccoy2014two} for a comprehensive overview). 
It is, therefore, possible to derive analytical expressions for the grand partition function and both the first and second moments of the \emph{relative} occupancy in thermodynamic equilibrium by employing the transfer matrix formalism. 
By taking the derivative of the grand partition function $\Xi$ with respect to $\mu$, we can calculate the mean \emph{relative} occupancy at equilibrium, $\langle\varphi\rangle = L^{-1} \sum_{i=1}^L\langle\varphi_i\rangle 
 =(L \Xi)^{-1}\partial_\mu \Xi $:
\begin{equation}
\langle \varphi \rangle = \frac{1}{2}\left[1 + \tanh{\left(\frac{L}{2{\xi}} \right)} \frac{\sinh{X}}{{\sqrt{\sinh^2{X} + \e^{-J}}}} \right],
\label{eq:phi}
\end{equation}
where $X = \left(J+\mu\right)/2$, and $\xi$ is the equilibrium correlation length of the system (given in number of lattice sites), which  can be rewritten in terms of $J$ and $\mu$ as follows (we refer to Section \ref{sec:SRLG} for more details):
\begin{equation}\label{eq:xi}
\xi = 1/\ln{\left(\frac{\cosh{X}+\sqrt{\sinh^2{X}+\e^{-J}}}{\cosh{X}-\sqrt{\sinh^2{X}+\e^{-J}}}\right)}.
\end{equation}
The mean equilibrium occupancy equals \mbox{$\langle N\rangle = L\langle\varphi\rangle$}. 
For purposes of illustration in what follows, unless otherwise stated, we shall use $L=13$. 
This particular choice will allow us to compare with greater ease the theoretical results presented here to the experimental data when we later, in Section \ref{sec:data}, apply our model to the study of the recruitment of torque-generating units (or stator units) onto the BFM.
It should be kept in mind that our theoretical results are valid for any value of $L$ and that our conclusions for small systems apply to any sufficiently small value of $L$.

We propose that the operational definition of ``small system'' be simply those systems for which fluctuations, as measured by the standard deviation of relative occupancy, are within experimental resolution and can be used to extract the sign and amplitude of the interaction strength within the usual biophysical range. 
A more careful rendering of this definition can be found in Section~\ref{sec:syssize}.
Such a definition is not contingent on the amplitude of (anti)cooperativity and therefore can encompass even the non-interacting (Hill-Langmuir) model.

\begin{figure}
\centering
\begin{subfigure}{0.49\textwidth}
\includegraphics[width=0.9\textwidth]{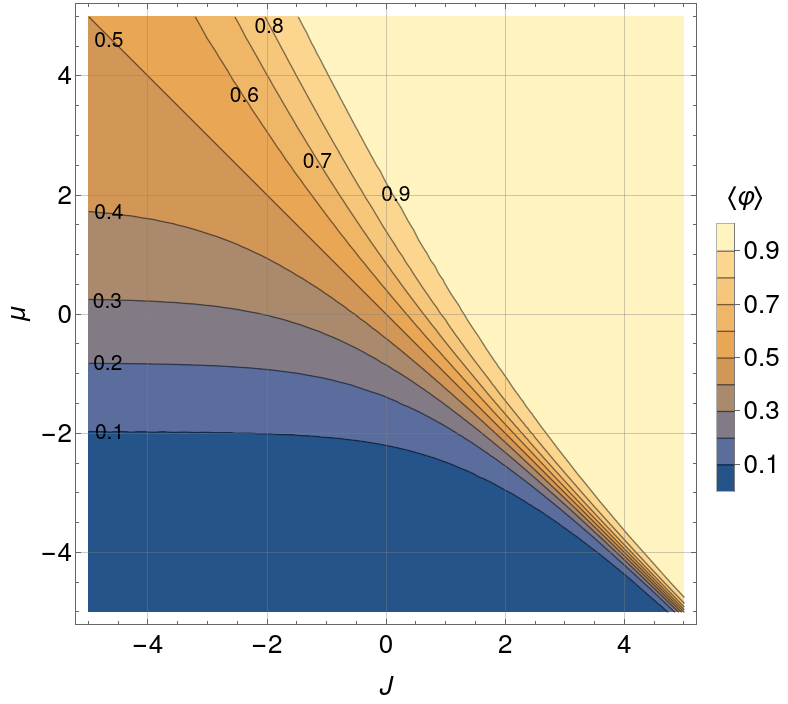}
\caption{}
\label{fig:Phi}
\end{subfigure}
\begin{subfigure}{0.49\textwidth}
\includegraphics[width=0.92\textwidth]{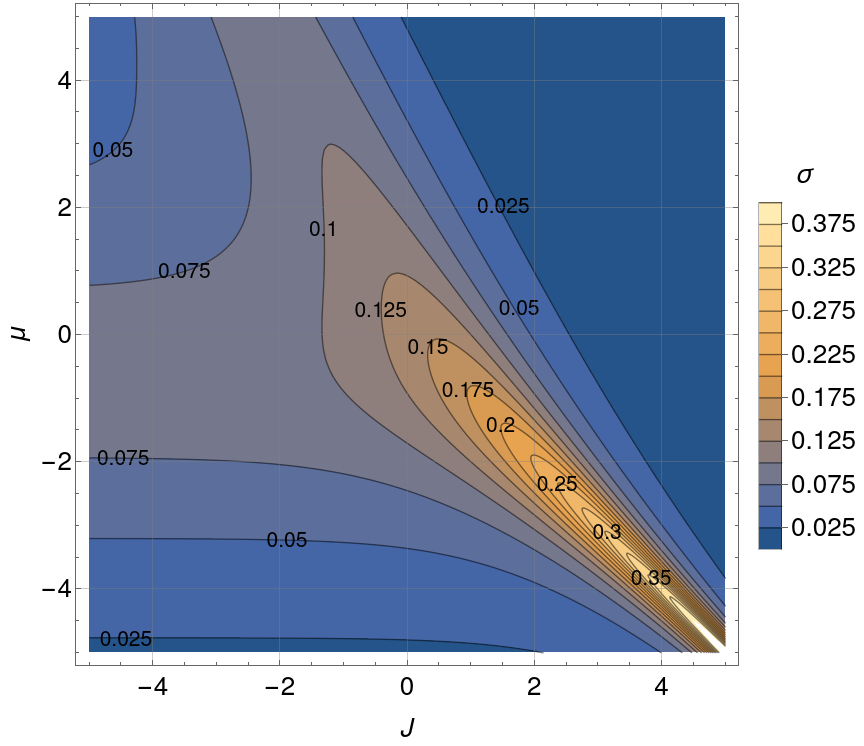}
\caption{}
\label{fig:SigmaPhi}
\end{subfigure}
\caption{Contour plots of the mean relative occupancy in equilibrium $\langle\varphi\rangle$ (left), and the standard deviation of the relative occupancy in equilibrium $\sigma = \sqrt{\langle \varphi^2 \rangle - \langle \varphi \rangle^2}$ (right), as functions of the dimensionless interaction potential $J$ and chemical potential $\mu$, according to expressions from Eq.~(\ref{eq:phi}) and (\ref{eq:sigmac}), for a system of size $L=13$.}
\label{fig:Phi+SigmaPhi}
\end{figure}

Fig.~\ref{fig:Phi} depicts the mean relative occupancy from Eq.~(\ref{eq:phi}) as a function of the two control parameters of the system: $J$ and $\mu$.
The half-filling (HF) contour line $\langle\varphi \rangle=0.5$ cuts the plot diagonally, meaning that the lattice is, on average, half-filled  whenever $J=-\mu$ (or $X=0$) holds.
Below (above) that line, the system's mean occupancy is 
lower (higher) than 50~\%.
Contour lines run together close to the plot's bottom right sector diagonal, corresponding to positive $J$ and negative $\mu$. 
In this sector defined by $X\approx 0$, for $J \gg 1$, the mean occupancy for a fixed positive value of the interaction potential becomes extremely sensitive to slight chemical potential variations.
Systems characterized by parameters in that range behave like biological switches, nearly jumping from zero to full occupancy with small changes in $\mu$.
On the contrary, more adaptive systems, that is, systems whose occupancy varies more smoothly from zero to full occupancy as a function of $\mu$, would be located in the central and upper left sector of  Fig.~\ref{fig:Phi}.

For the short-range lattice gas model from Eq. (\ref{eq:Hamiltonian}), the standard deviation $\sigma$ of the occupancy at equilibrium can be expressed in terms of the derivative of the mean relative occupancy
with respect to the chemical potential as follows (cf. Section \ref{sec:SRLG}): 
$\sigma = \sqrt{\langle \varphi^2 \rangle - \langle \varphi \rangle^2} 
 =  L^{-1/2} \sqrt{ \partial_\mu \langle \varphi \rangle}$.
 Applying this result to Eq.~(\ref{eq:phi}), we obtain a
 compact
 analytical expression for $\sigma$:
\begin{equation}  \label{eq:sigmac}
      \sigma   =  \frac{1}{2}
\left[
 \sech^2 \left( \frac{L}{2\xi}\right) \frac{\sinh^2{{X}}}{\sinh^2{{X}}+\e^{-J}}
+ \tanh{\left(\frac{L}{2{\xi}} \right)}
\frac{\e^{-J} \cosh{{X}}}{L\left(\sinh^2{{X}} + \e^{-J}\right)^{3/2}}
\right]^{1/2},
\end{equation}
whose dependency on $J$ and $\mu$ is depicted in Fig.~\ref{fig:SigmaPhi}. 
In what follows, we will use $\sigma$ as a measure of fluctuations, hence the stochastic nature of the system. 

\begin{figure}
    \centering
    \includegraphics[scale=0.35]{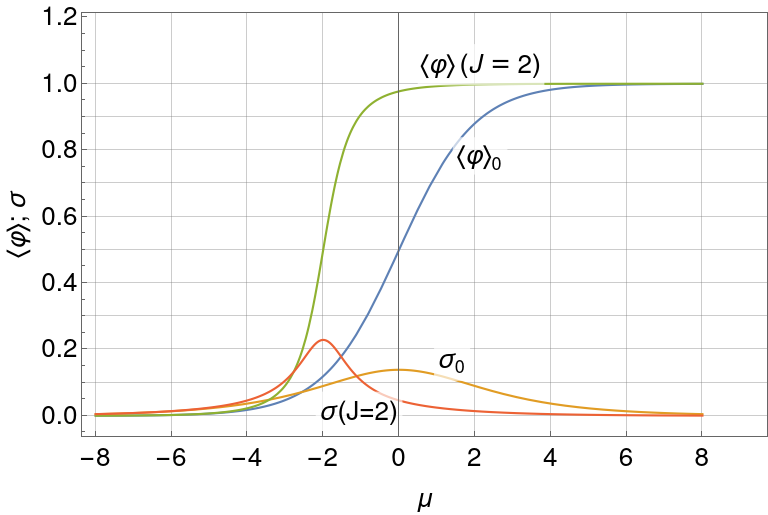}
    \caption{Comparison between the average relative occupancy and the standard deviation as functions of  $\mu$ for two values of cooperativity: $J = 0$ (Hill-Langmuir) and 
    $J = 2$.}
    \label{fig:phi0_vs_phi2}
\end{figure}

In Fig.~\ref{fig:phi0_vs_phi2}, we compare the average relative occupancy, $\langle \varphi \rangle$, and the standard deviation, $\sigma$, as functions of  $\mu$ for two vertical cuts in the above contour plots: $J = 0$ (Hill-Langmuir) and $J = 2$. 
We note that with increasing $J$ the $\langle \varphi \rangle$ curve sharpens and shifts to lower values of $\mu$  and that the maximum in $\sigma$ increases in amplitude and also shifts to lower values of $\mu$. 
These shifts are due to the displacement with increasing $J$ of the value of $\mu$ at half-filling, $\mu_{\rm HF}(J) = - J$, where the slope of the $\langle \varphi \rangle$ curve and the amplitude of the standard deviation are both largest.

\subsubsection{System behavior at half-filling}
Fluctuations increase as $J > 0 $ grows. Along the half-filling diagonal ($X=0$)  in Fig.~\ref{fig:SigmaPhi}, they saturate at a finite global maximum as $J$ goes to infinity.

By evaluating Eq. (\ref{eq:sigmac}) and (\ref{eq:xi}) for $X=0$, one obtains an explicit formula for the standard deviation at half-filling (HF) as a function of $L$ and $J$:
\begin{equation}
\label{eq:sigmaHF}
\sigma_\text{HF}\left(J\right)=\frac{\text \e^{J/4}}{2\sqrt{L}}\sqrt{\text{tanh}\left(\frac{L}{2\xi_\text{HF}\left(J\right)}\right)},
\end{equation}
with
\begin{equation}\label{eq:xiHF}
	\xi_{\rm HF} = 
	1/\ln{\left( \frac{1+\text \e^{-J/2}}{1-\text \e^{-J/2}}\right)} = 1/\ln [\coth (J/4)],
\end{equation}
which represents the maximum equilibrium correlation length at fixed $J$ for any value of mean occupancy. 

In the Hill-Langmuir limit ($J=0$), the correlation length vanishes, and the standard deviation at HF becomes $1/(2\sqrt{L})$. 
For low couplings ($0 \le J <2$), the correlation length increases sub-linearly with $J$ and is accurately given by $\xi_{\rm HF} \approx -1/\ln(J/4)$.  In this same limit, the standard deviation increases exponentially with $J$,
$\sigma_\text{HF}(J)/\sigma_\text{HF}(0) \approx \e^{J/4}$, 
an approximation valid for $J<4$, when $L \approx 10$. This increase can be interpreted as a widening of the probability distribution function of states of a given $N$ around the average value $\langle N \rangle$, where the probability is a maximum.
In the high cooperativity limit $J \gg 1$, $\xi_{\rm HF}$ shows an asymptotic exponential growth, $\xi_{\rm HF} \sim \frac{1}{2}\e^{J/2}$, which is the analog of the well-known zero temperature critical behavior of spin chains \cite{Godreche2000response}; and, consequently, for fixed $L$, $\sigma_\text{HF}$ tends to 1/2 when the correlation length surpasses $L/2$ and the system evolves into a bimodal one (for which only the zero and full occupancy states contribute, see below).
We remark that although for fixed $J$, $\sigma_\text{HF}$ vanishes as $L^{-1/2}$ when $L\to\infty$, as expected, in the limit where $L$ is kept fixed and $J$ tends to $+\infty$, the standard deviation at HF reaches the finite (maximum, or saturation) value of one-half.

Fluctuations decrease as $J < 0 $ becomes more negative. Along the half-filling diagonal ($X=0$) in Fig.~\ref{fig:SigmaPhi} ($L=13$), they saturate at a finite value as $J \to - \infty$. 
This is a general result for systems with an odd number of sites.
In this strong anti-cooperative limit, for small enough systems, it turns out that fluctuations depend sensitively on whether the number of lattice sites is even or odd. 
For $L$ even, $\sigma_\text{HF}$ tends to zero; for odd $L$, it tends to $ 1/(2L)$. 
This leads to the striking result that in the limit of strong anti-cooperativity, a fluctuation analysis near HF could clearly detect the difference between small systems with even and odd numbers of sites, provided that the experimental precision is high enough.

\subsection{\texorpdfstring{Fluctuations of the mean occupancy and their dependency on $\braket{\varphi}$ and $J$}{Fluctuations of the mean occupancy and their dependency on <φ> and J}}

Often, the experimental parameter one can measure (or control indirectly, as is the case for the BFM 
by varying the external load, 
cf. Section~\ref{sec:data}) is not the chemical potential of the reservoir $\mu$, but the mean number of bound ligands at equilibrium, $\langle N \rangle$. 
It is, therefore, insightful to plot the standard deviation $\sigma$ as a function of the mean relative occupancy $\langle \varphi \rangle$, which, in general, can be done using a parametric plot.

\begin{figure}
\centering
\begin{subfigure}{0.45\textwidth}
\includegraphics[width=0.9\textwidth]{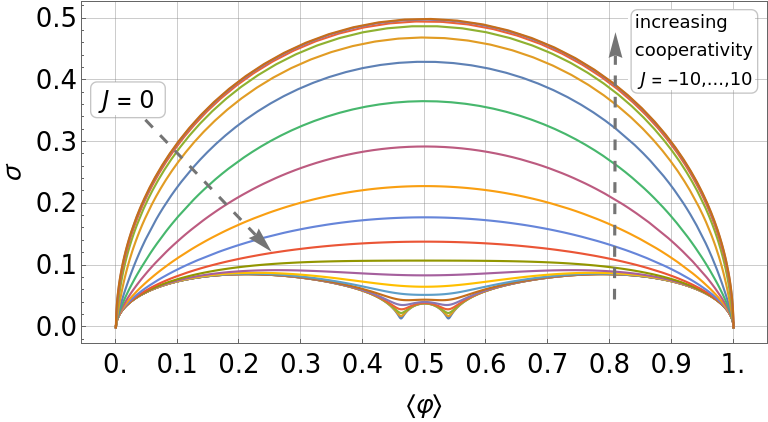}
\caption{\label{fig:SigmaPhiVsPhi}}
\end{subfigure}
\begin{subfigure}{0.45\textwidth}
\includegraphics[width=0.9\textwidth]{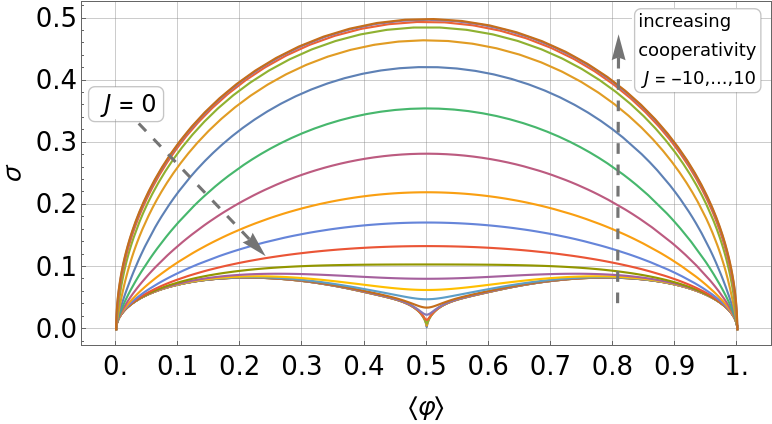}
\caption{}
\label{fig:SigmaPhiVsPhi_L14}
\end{subfigure}
\caption{\label{fig:SigmaPhiVsPhi+L13+L14} Parametric plots of the standard deviation of the relative occupancy, $\sigma$, versus the mean relative occupancy, $\langle\varphi\rangle$, for different values of the interaction potential: from $J=-10$ (bottom curve) to $J=10$ (upper curve) in steps of $1$. Plots (a) and (b) correspond to systems of size $L=13$ and $L=14$, respectively.}
\end{figure}
   
To unfold this relation, we express $\langle \varphi\rangle$ and $\sigma$, for different but fixed values of $J$, as functions of $\mu$ and plot this family of curves in Fig.~\ref{fig:SigmaPhiVsPhi} and Fig.~\ref{fig:SigmaPhiVsPhi_L14} for an odd and even number of lattice sites, respectively.
Each curve for a given $J$ represents the set of points of coordinates $\left(\langle\varphi\rangle=f_J(\mu), \,\sigma=g_J(\mu) \right)$ with $\mu$ spanning from $-15$ to $30$, which turns out to be a suitable one-dimensional domain to complete the plots. 
We have chosen the interaction potential $J$ to take values from $-10$ to $10$ in steps of size 1. 
The standard deviation shows a left-right \emph{particle-hole} symmetry with respect to the system half-filling: $\langle\varphi\rangle_\text{HF}= f_J(-J)=1/2$, which is a manifestation of the underlining particle-hole symmetry of the system under study. 
The system has zero fluctuations when $\langle\varphi\rangle=0$ (empty substrate) or $\langle\varphi\rangle=1$ (full substrate), which is easy to understand because a distribution of states with non-zero probability around these average values would be incompatible with the average values themselves.
For non-negative values of $J$, the maxima of fluctuations appear at half-filling, taking on the value of $1/(2 \sqrt{L})$ for $J=0$ and 1/2 for $J\to +\infty$, and the $\sigma$ vs. $\langle\varphi\rangle$ curves converge asymptotically towards the limiting $J\to +\infty$ one.
More subtly, results for the anti-cooperative limit $J\to -\infty$ depend on whether $L$ is odd or even.

A key result of the above analysis and one of our major conclusions is that for cooperative systems of size $L\approx 10$ and mean relative occupancies not too close to 0 or 1,  in the range $0<J<7$ the curves of $\sigma$ vs. $\langle\varphi\rangle$ are spaced sufficiently far enough apart that the standard deviation is suited for estimating $J$. 
The same conclusion can be drawn for anti-cooperative systems of size $L\approx 10$ and mean relative occupancies close to 1/2 in the range $-3<J<0$.
These results are illustrated in Fig.~\ref{fig:SigmaPhiVsPhi} for $L=13$ and Fig.~\ref{fig:SigmaPhiVsPhi_L14} for $L=14$. 
Outside these ranges, the standard deviation saturates to limiting values, and therefore experimental data near these limits could only be used to put bounds on the value of $J$.

We will investigate the cooperative and anti-cooperative regimes in more detail in what follows, taking care to study those limits for which we were able to derive simplified exact or approximate analytical expressions for $\sigma(\langle\varphi\rangle)$.

\subsubsection{Fluctuations in the zero-cooperativity limit: the Hill-Langmuir case}
In the absence of cooperativity ($J=0$), which corresponds to the case of a simple adsorption process with on-site volume exclusion interactions only, the free energy calculated from Eq.~(\ref{eq:Hamiltonian}) reduces to an expression determined entirely by the mesostate (the occupancy) $N$: $\beta\mathcal{H}_0(\bm{\varphi} | N)=-\mu N$. 
The correlation length from Eq. (\ref{eq:xi}) vanishes; hence each adsorbing site acts as an independent particle trap.
Thus we recover the Hill-Langmuir adsorption model at equilibrium for which $\langle \varphi \rangle$ takes the following typical sigmoid form as a function of $\mu$:
\begin{equation}
\langle \varphi \rangle_0(\mu) = 1/\left(1+\e^{-\mu}\right).
\end{equation}
The corresponding standard deviation becomes:
\begin{equation}\label{eq:sigma0}
    \sigma_0(\mu) =  \frac{\sech\left(\mu/2\right)}{2\sqrt{L}}=\sqrt{\frac{\langle\varphi\rangle_0 - \langle\varphi\rangle_0^2}{L}},
\end{equation}
which can be reinterpreted to obtain
\begin{equation}\label{eq:sigma0phi}
    \sigma_0(\langle\varphi\rangle) =  \sqrt{\frac{\langle\varphi\rangle - \langle\varphi\rangle^2}{L}}.
\end{equation}
This latter expression exemplifies the general result that, in the thermodynamic limit, $L\to\infty$, $\langle \varphi^2 \rangle \to \langle \varphi \rangle^2$, and hence, as expected, the standard deviation (of the \emph{relative} occupancy) tends to zero as $L^{-1/2}$ 
(self-averaging property at equilibrium in the thermodynamic limit). 

In Section \ref{sec:SM-pdf}, we show 
using a Gaussian approximation for the PDF 
that for $J=0$ the standard deviation is inversely proportional to the curvature (in absolute value) of the entropy of mixing, which reaches a minimum at HF. Therefore, the standard deviation for $J=0$ reaches a maximum there. More generally, one can show that for $J>0$ not too large, a Gaussian approximation for the PDF is accurate and therefore the standard deviation is inversely proportional to the curvature (in absolute value) of an effective free energy, which reaches a minimum at HF.

\subsubsection{Fluctuations in the strong cooperativity limit}
Away from HF, although $\xi$ remains finite in the large $J$ limit, it can still be large compared with $L/2$ for not-too-large values of $L$, provided that one is not too close to zero or full filling (see Fig.~\ref{fig:xiJinf}). 
Taking the limits $J \to \infty$  and $\mu = 2 X - J \to -\infty$  simultaneously, but keeping $L$ finite and $X$ constant, allows Eqs.~\eqref{eq:phi} to be simplified to 
\begin{equation}
	\langle \varphi \rangle_\infty(X) = 
    \frac{1}{2}\left[1 + \tanh{\left(\frac{L}{2{\xi_\infty(X)}} \right)} \text{sgn}(X)\right],  
    \label{eq:phiJinf}
\end{equation}
and $\langle \varphi^2 \rangle_\infty = \langle \varphi \rangle_\infty$, where $\xi_\infty(X) = 1/(2 |X|)$, leading to an infinite $J$ standard deviation of
\begin{equation}\label{eq:sigmaInf}
\sigma_\infty(X) 
= \frac{1}{2} \sech \left( \frac{L}{2\xi_\infty(X)}\right)
= \sqrt{\langle \varphi \rangle_\infty - \langle \varphi \rangle_\infty^2},
\end{equation}
which can be reinterpreted to obtain
\begin{equation}\label{eq:sigmaInfphi}
    \sigma_\infty(\langle\varphi\rangle) =  
    \sqrt{\langle \varphi \rangle - \langle \varphi \rangle^2}.
\end{equation}
The maximal standard deviation at half-filling is due to the large width of the distribution of states with non-zero probability compatible with $\langle\varphi\rangle = 1/2$ (see Section \ref{sec:ProbDist}). 
This limit is easy to understand in terms of a bimodal system:  only the empty and full states contribute because the probability of intermediate states is suppressed by the presence of energetically costly domain walls (as will be discussed in detail below). 
The bimodal system, therefore, behaves as an effective Hill-Langmuir model with a system size equal to 1, as can be seen by taking $L=1$ in Eqs.~\eqref{eq:sigma0phi}.

\subsubsection{Fluctuations for intermediate finite cooperativity} 

\begin{figure}[ht]
\centering
	\includegraphics[width=86mm]{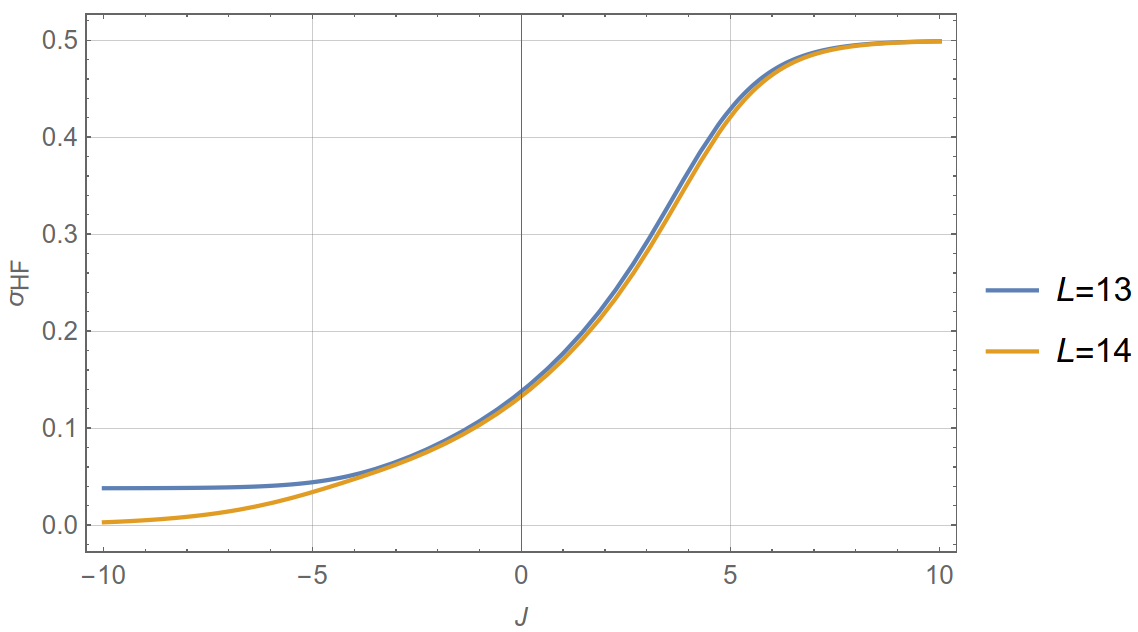}
	\caption{\label{fig:SigmaPhiHF} Standard deviation of the mean occupancy at half-filling, $\langle \varphi \rangle=1/2$, as a function of the dimensionless interaction potential $J$ for a system of size $L=13$ (blue) and $L=14$ (orange).
    For $J \ge 0$ the prediction of the block domain approximation, Eq.~(\ref{eq:sigmablock}),  is indistinguishable from the exact result.}
\end{figure}

We have seen in Fig.~\ref{fig:SigmaPhiVsPhi+L13+L14} how, at fixed average occupancy, increasing cooperativity leads to an increase in fluctuations.  
A simple, insightful, and accurate way of understanding this increase can be formulated by starting from the previously derived result for the Hill-Langmuir model and introducing the concept of block domains. 
When $J=0$, all sites are decoupled, and the standard deviation is non-zero for a finite-size system simply because of statistical fluctuations (absence of self-averaging).  
When $J>0$, neighboring sites are coupled and can be grouped (approximately) into correlated block domains of size $b(\xi, L) > 1$, leading to 
\be
\label{eq:sigmablock}
\sigma(J; \xi, L) \approx  L_{\rm eff}^{-1/2} \sqrt{\langle\varphi\rangle - \langle\varphi\rangle^2},
\ee
where $L_{\rm eff} \equiv L/b$.   
When $1 < 2 \xi < L$, correlations lead to block domains that we take to be of size $b = 2 \xi$.
The above approximation allows the system to be described by a reduced number, $L_{\rm eff}$,  of independently fluctuating block domains of size $b$.
Since  $b$ should also tend to 1, when $\xi \to 0$ (zero cooperativity, Hill-Langmuir); and to $L$, when $2\xi/L \gg 1$ (strong cooperativity, bimodal behavior), a convenient interpolation formula between the above three cases is 
\be
b(\xi, L) = \frac{\tanh\left[L/(2\xi)\right]}{\tanh\left[1/(2\xi)\right]}.
\ee
By comparing the predictions of this approximation with the exact results at half-filling, where the standard deviation is a maximum, one can see that this approximation is extremely accurate (see Fig.~\ref{fig:SigmaPhiHF}). 

Exact results for the standard deviation can be obtained  by writing the  mean square relative occupancy at equilibrium, 
\be
 \langle \varphi^2 \rangle =
 \frac{1}{L^2  \Xi}\partial^2_\mu \Xi =
 L^{-2} \sum_{i=1}^L \sum_{j=1}^L\   
 \langle  \varphi_i \varphi_j \rangle,
 \label{eq:phi2sum}
 \ee
in terms of the 2-point correlation function, $C_{ij} =\langle  \varphi_i \varphi_j \rangle$ 
(with $C_{ii} = \langle  \varphi_i^2 \rangle =  \langle  \varphi_i \rangle = \langle  \varphi \rangle$).  

Eq.~(\ref{eq:phi2sum}) can be used as a starting point to gain deeper physical insight into the behavior of the standard deviation $\sigma$ than the previous result obtained directly from the grand partition function because it becomes possible to express $\sigma$ as an explicit function of $\langle  \varphi \rangle$ and the correlation length $\xi$, see Eq.~(\ref{eq:sigmacorr}).

For $J>0$, the standard deviation is bounded by the zero and infinite cooperativity values, which can be calculated from the limiting forms of  $C_{ij}$.
In the absence of cooperativity ($J=0$), $C_{ij}$ factorizes for $i\neq j$:
$C_{ij} \rightarrow \langle \varphi_i \rangle  \langle \varphi_j \rangle = \langle \varphi \rangle^2 $. 
In this $J=0$ limit,  $\langle  \varphi^2 \rangle$ can then be obtained directly from Eq.~\eqref{eq:phi2sum} and  $\sigma$
simplifies to  
$\sigma_0$ (Eq.~(\ref{eq:sigma0phi})).
In the other limit of infinite cooperativity ($J\rightarrow \infty$) and strong correlations, $\xi$ as a function of 
$\langle \varphi \rangle$
diverges at HF and away from HF for not too large $L$ remains larger than $L/2$ 
(see Fig.~\ref{fig:xiJinf}). 
From Eq.~(\ref{eq:corrfct}) we see that for $L$ finite in this limit $C_{ij} \rightarrow \langle \varphi_i^2 \rangle = \langle \varphi \rangle$,  and $\sigma$ saturates at $\sigma_\infty$  (Eq.~(\ref{eq:sigmaInfphi})).

\subsubsection{Fluctuations in the strong anti-cooperativity limit}
By comparing the $L=13$ and 14 cases (Figs.~\ref{fig:SigmaPhiVsPhi} and \ref{fig:SigmaPhiVsPhi_L14}), we observe that for small enough anti-cooperative  systems, there is a clear difference between even and odd system sizes. 
This difference clearly manifests itself near HF, and, as we will see below, this difference arises because of frustration for odd $L$.

For negative values of $J$ (anti-cooperativity) \emph{and} $L$ even the curves inflect and tend for $J \ll -1$ towards a characteristic limiting shape with global maxima near $\langle\varphi\rangle = 0.2$ and 0.8 and a zero at half-filling (see
Fig.~\ref{fig:sigmaeven}).
A  restricted grand partition function approach (presented in Section \ref{sec:SM-pdf}), which
retains only non-overlapping particle-hole pairs, leads to a simple but accurate approximation in the limit $J\to -\infty$ for $L$ even
(see Fig.~\ref{fig:sigmaeven}):
\begin{equation}
\sigma^{\rm even}_{-\infty} = L^{-1/2} \sqrt{\langle \varphi \rangle - 2 \langle \varphi \rangle^2} \qquad\text{for}\quad 0 \le \langle \varphi \rangle \le 1/2.
\end{equation}
The corresponding standard deviation for $1/2 \le \langle \varphi \rangle \le 1$ can be obtained by exploiting particle-hole symmetry: $\sigma(\langle \varphi \rangle) = \sigma(1-\langle \varphi \rangle)$.

In this limit of strong anti-cooperativity ($J\to -\infty$), the correlation length becomes complex, $1/\xi \simeq i \pi + 2 \cosh (X) \e^{-|J|/2} $. 
Hence, at HF, $1/\xi \simeq i \pi$, leading directly to an oscillating two-point correlation function that describes a sequence of non-overlapping particle-hole pairs (anti-ferromagnetic order in Ising spin language):
\begin{equation}
C_{i,i+r}^{\rm HF} \approx  \frac{1 + \cos (\pi r)}{4}\qquad \text{for}\quad J\to-\infty,
\end{equation}
where we have taken the real part of $C_{i,i+r}$ and assumed that $r/L \ll 1$. 
$C_{i,i+r}^{\rm HF}$ alternates between 0 and 1/2, which reflects alternating perfect anti-correlations (nearest neighbors) and correlations (next nearest neighbors) at HF in this limit of strong anti-cooperativity.

\subsection{\texorpdfstring{Probability distribution function (PDF) of the occupancy and its dependency on $\langle \varphi \rangle$ and $J$}{Probability distribution of the occupancy and its dependency on <φ> and J}\label{sec:ProbDist}}

We can extract the equilibrium probability distribution function (PDF) of the occupancy (corresponding to the effective Hamiltonian given in Eq.~(\ref{eq:Hamiltonian})) for fixed $J$ and $\mu$, $P(N; J, \mu)$ through exact enumeration and study how it depends on the interaction potential $J$. 
To study $P(N; J, \lan \varphi \ran)$ for fixed $J$ and $\lan \varphi \ran$, the experimentally accessible quantity, we varied $J$ but kept the mean occupancy fixed  (by adjusting $\mu$ to any given choice of $J$).

\begin{figure}
\centering
\begin{subfigure}{0.32\textwidth}
\includegraphics[width=0.95\textwidth]{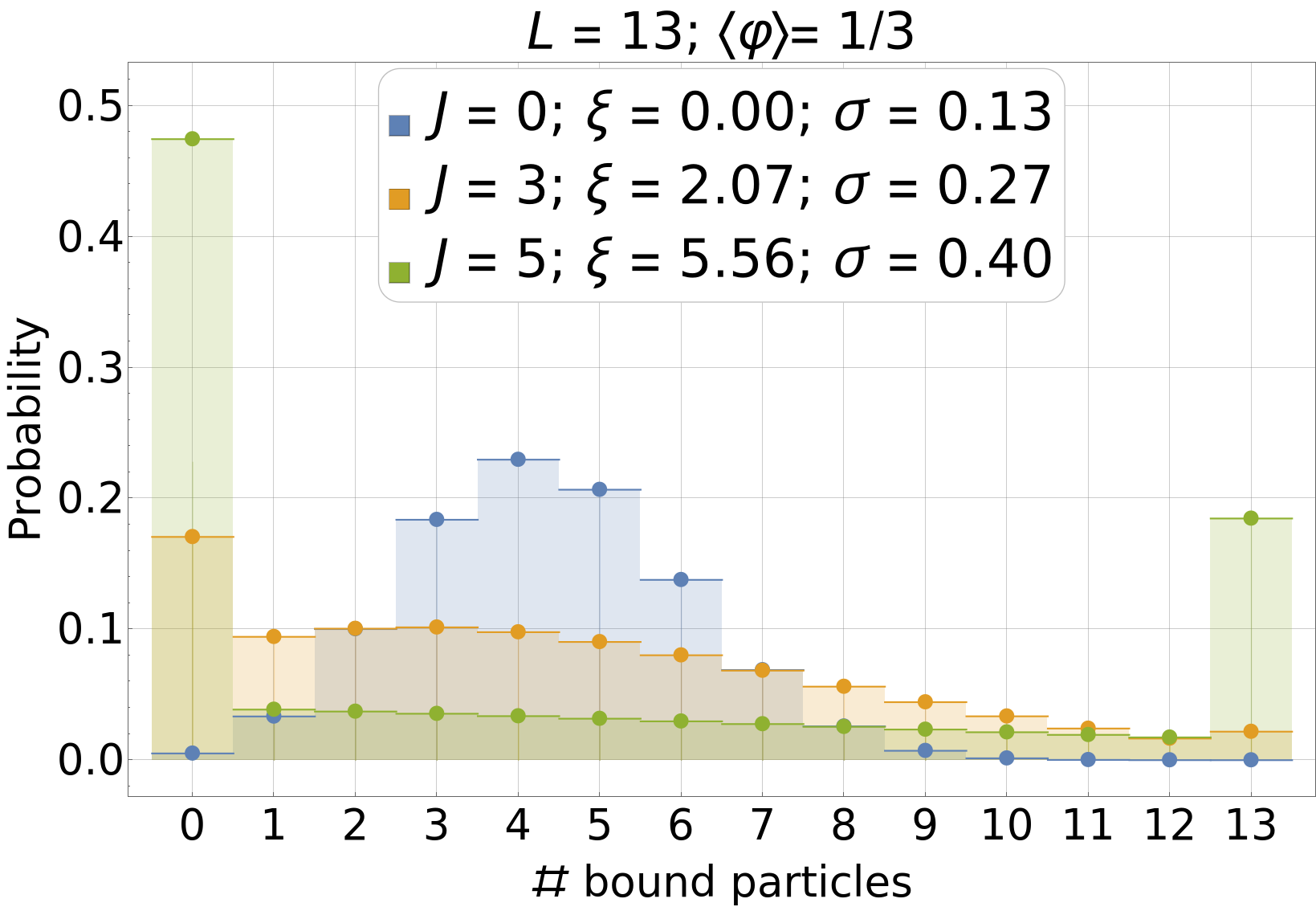}
\caption{}\label{fig:OccupanyPDFa}
\end{subfigure}
\begin{subfigure}{0.32\textwidth}
\includegraphics[width=0.95\textwidth]{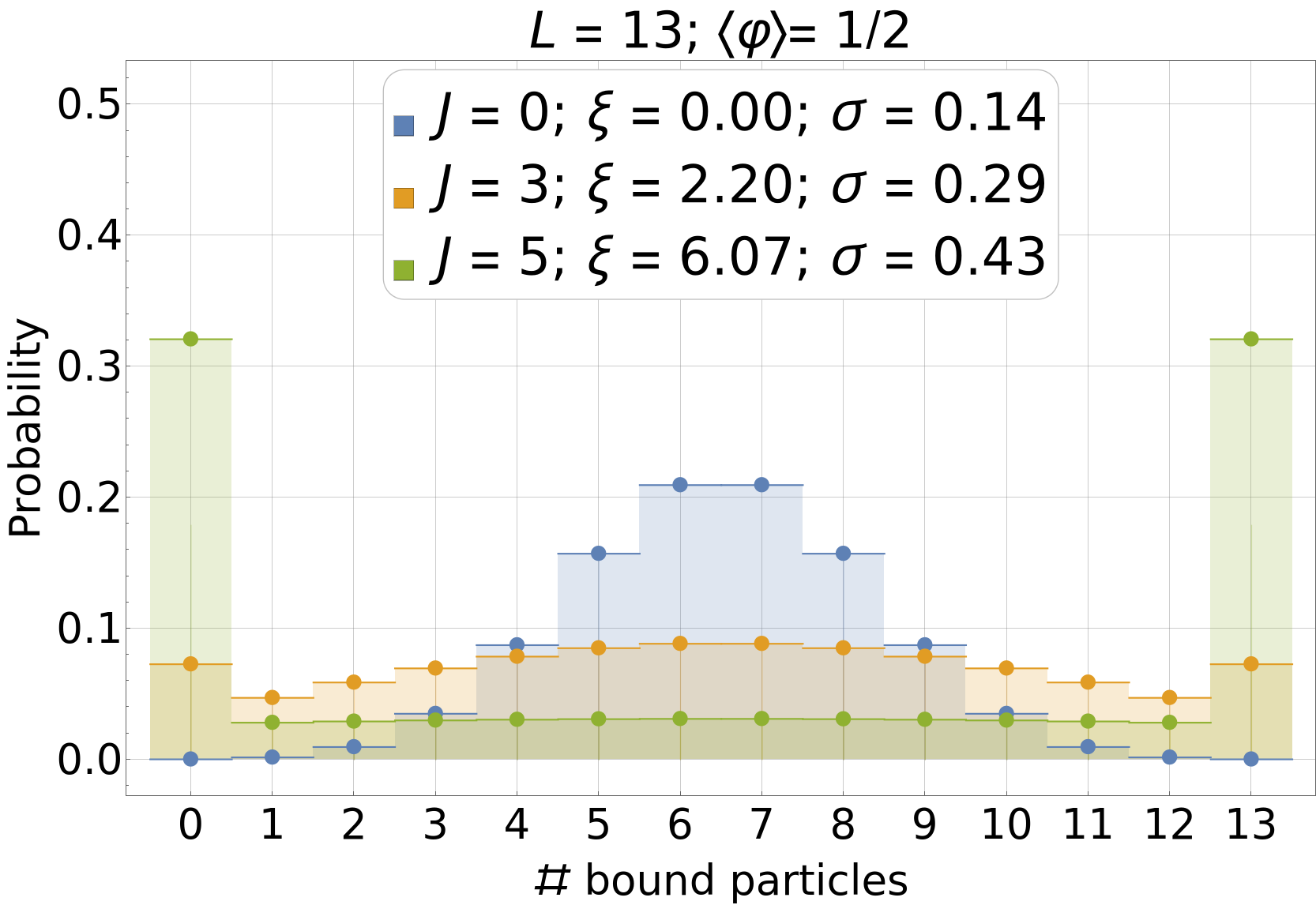}
\caption{}\label{fig:OccupanyPDFb}
\end{subfigure}
\begin{subfigure}{0.32\textwidth}
\includegraphics[width=0.95\textwidth]{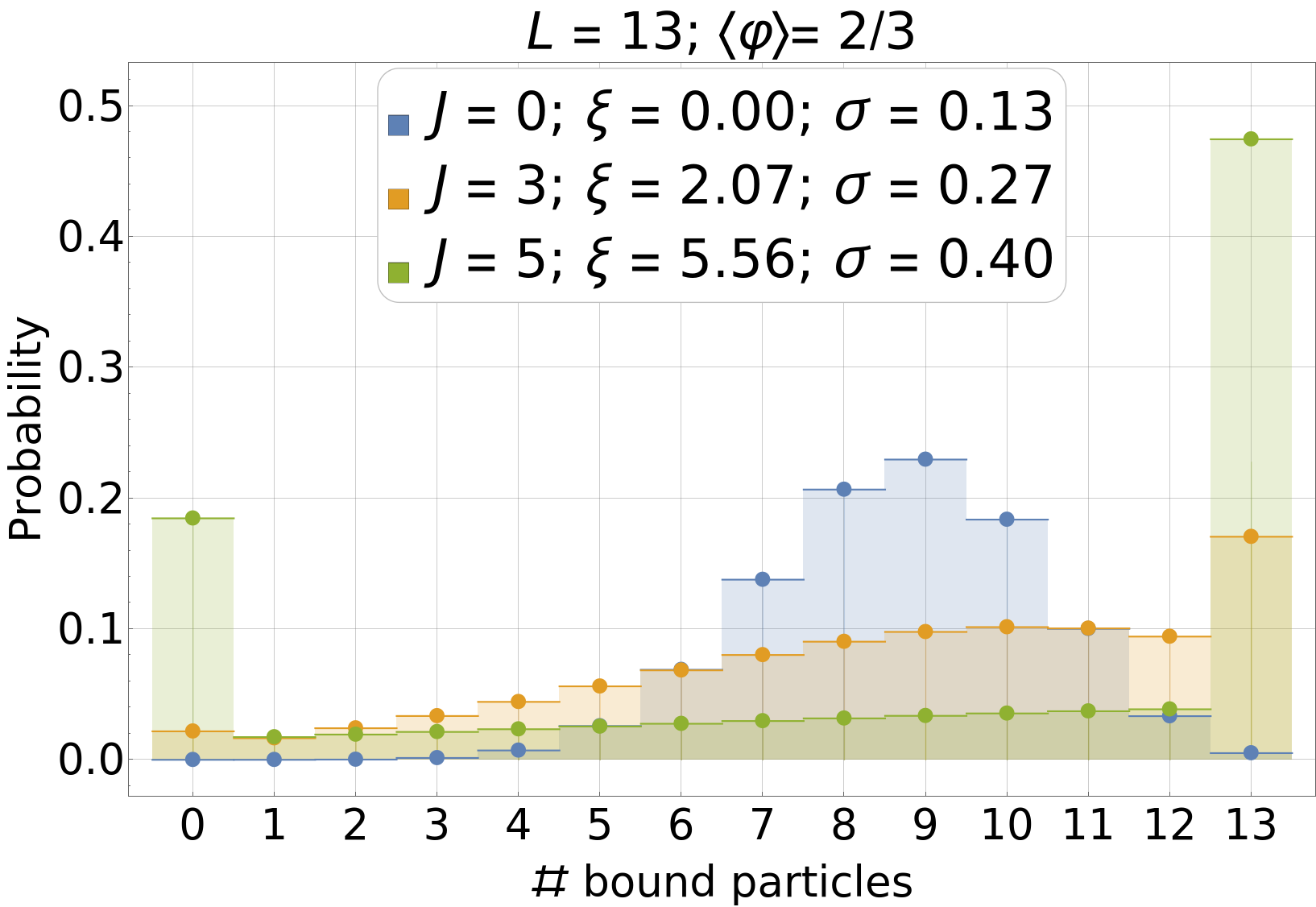}
\caption{}\label{fig:OccupanyPDFc}
\end{subfigure}
\caption{Discrete equilibrium probability distribution of the number of bound particles, $P\left(N\right)$, for different values of the (dimensionless) interaction potential: $J=0$ (orange), 3 (blue), and $5$ (green). Histograms are obtained from an exact enumeration using the 1D Hamiltonian from Eq.~(\ref{eq:Hamiltonian}) for a system of size $L=13$ for the average occupancies $\langle N \rangle = \langle\varphi\rangle L = 4.33,\,6.5,$ and $8.67$ ($\langle\varphi \rangle = 1/3,\, 1/2,$ and $2/3$), respectively.}
\label{fig:OccupanyPDF}
\end{figure}

Taking $L = 13$, Fig.~\ref{fig:OccupanyPDF} shows how the PDF behaves for three different positive values of the interaction potential and three characteristic values of the mean relative occupancy: $\langle\varphi\rangle = 1/3,\,1/2,$ and $2/3$, respectively.
The correlation length increases with $J$, and so does the standard deviation; consequently, the PDF profile broadens and flattens. 
On the one hand, finite-size effects \emph{set in} when the tails of the PDF touch the system boundaries.
On the other hand, when the typical size of a highly correlated particle domain, $\approx 2\xi$,  becomes comparable to the substrate size, $L$, finite-size effects \emph{dominate}, and the PDF saturates to its strong interaction bimodal form.

\subsubsection{PDF in the zero-cooperativity limit: the Hill-Langmuir case}
In the absence of cooperativity, we expect $N$ to be normally distributed with standard deviation $L \sigma_0$,\footnote{Strictly speaking, the variable $N$ being discrete, the probability follows a binomial distribution. But the latter is well approximated by the normal distribution in the large $L$ limit.} provided the system is large enough, and the position of the mean is far enough from the system boundaries:  $P_0\left(N; \langle\varphi\rangle \right) \equiv P\left(N; 0,\langle\varphi\rangle \right) \approx \mathcal{N}\left(\langle N\rangle, L\sigma_0\right)$ for $L\gg 1$.
The half-filling case with $J=0$ in Fig.~\ref{fig:OccupanyPDFb} approximates this Gaussian behavior well because the average value, $\langle N\rangle$, is far enough from the boundaries that the tails of the distribution do not reach the boundary values, 0 and $L$ (the distribution decays fast enough to the left and right of the average value). 
More quantitatively, for any value of $J$, finite-size effects set in when $\langle N\rangle <  2L \sigma $ or $L-\langle N\rangle <  2L \sigma$ (i.e., when the tails of the PDF touch the system boundaries). 

\subsubsection{PDF in the strong cooperativity limit}
The PDF broadens and flattens for intermediate positive values of the interaction potential.
For higher values, e.~g. $J\geq5$, the PDF saturates by accumulating at the boundaries at 
\begin{equation}
P_\infty\left(N; \langle\varphi\rangle \right) =  (1-\langle\varphi\rangle) \delta_{N,0}
+  \langle\varphi\rangle \delta_{N,L}
\end{equation}
(here, $\delta_{ij}$ is the Kronecker delta) 
and the system becomes well described by an effective two-state (bimodal) system (this is the limit where there exists only one effective block domain).  
This result is intuitively plausible and can be obtained by retaining in the full partition sum only the empty and full states (see Section \ref{sec:SM-pdf} for more details).
In this limit, finite-size effects are always dominant, and the standard deviation for the occupancy $N$ can easily be calculated directly from the PDF. 
As expected, it saturates at $L\sigma_\infty$  ($=L/2$ at half-filling). 
Owing to the simple form of this limiting PDF, it is easy to calculate all moments of $N$, leading to $\langle N^m \rangle_\infty = L^m \langle \varphi \rangle_\infty$ (or $\langle \varphi ^m \rangle_\infty = \langle \varphi \rangle_\infty$) and therefore explicit expressions for low order standardized moments, such as the skewness and kurtosis, see Section \ref{sec:SM-pdf}.
This simplification occurs because, with the correlation length, the energy cost needed to create domain walls also increases with $J$.
Hence, a strongly correlated system favors microscopic configurations that minimize the number of domain walls, selecting the mesostates corresponding to either an empty or a fully occupied substrate.
This tendency to select extremal occupancies is a characteristic signature of the presence of strong cooperativity effects in small-size systems.
Furthermore, for a parameter choice corresponding to an occupancy expectation value different from half-filling, the distribution is skewed, as in Fig.~\ref{fig:OccupanyPDFa} and ~\ref{fig:OccupanyPDFc}, with an asymmetry that increases with the value of the interaction potential.

\subsubsection{PDF in the strong anti-cooperativity limit}
\begin{figure}
\centering
\begin{subfigure}{0.45\textwidth}
\includegraphics[width=0.9\textwidth]{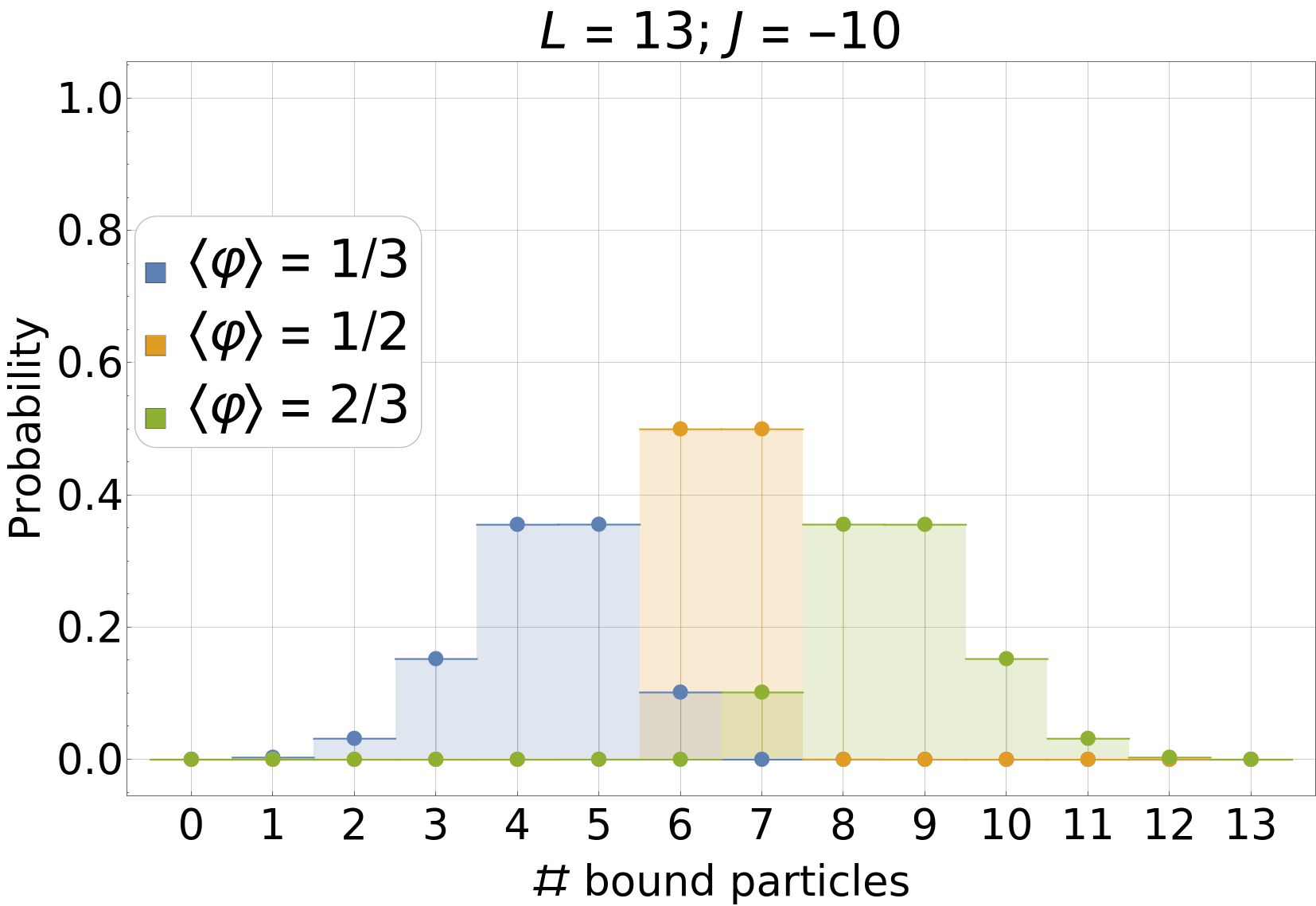}
\caption{}\label{fig:PDF-strong-anticooperqtivitya}
\end{subfigure}
\begin{subfigure}{0.45\textwidth}
\includegraphics[width=0.9\textwidth]{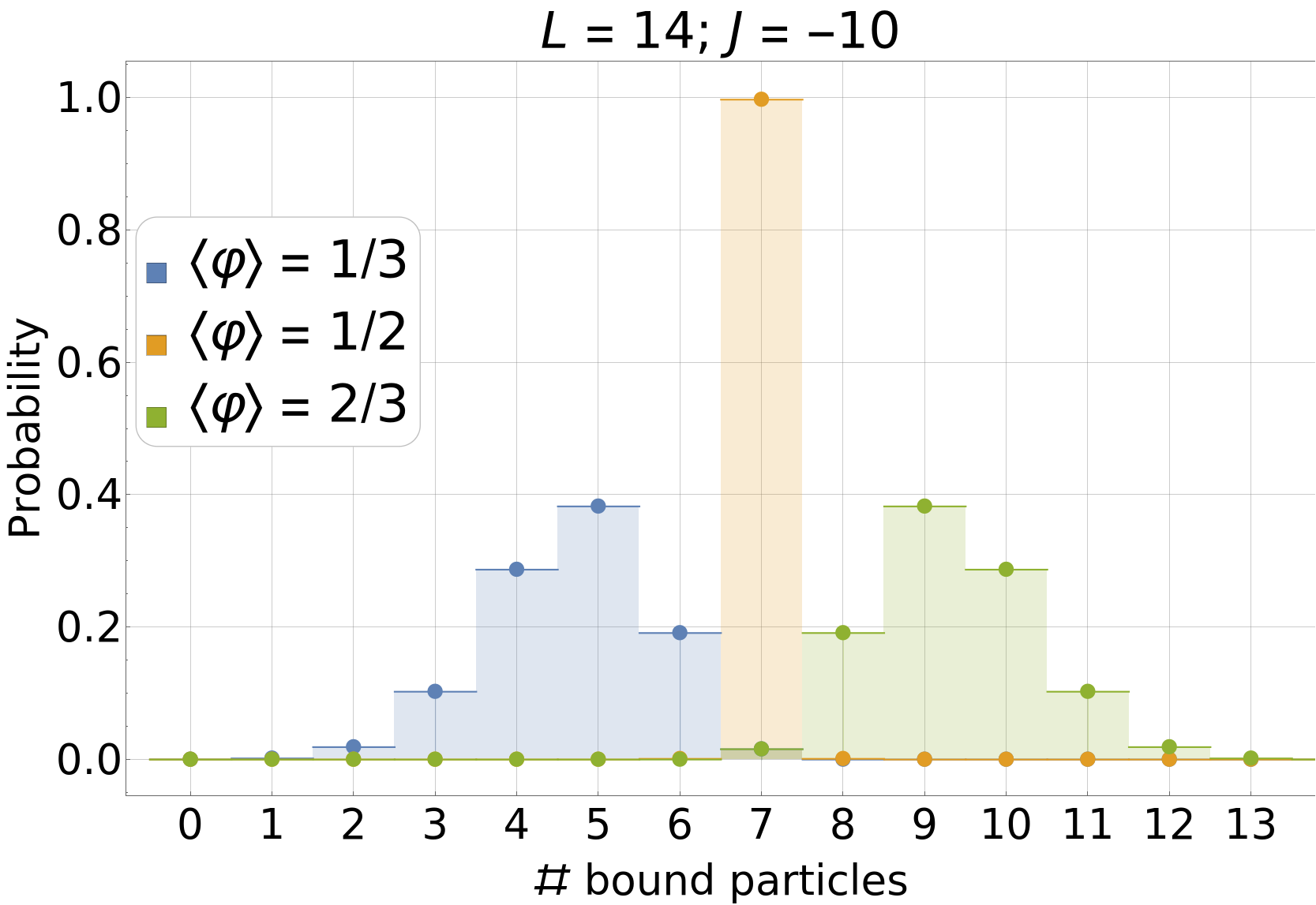}
\caption{}\label{fig:PDF-strong-anticooperqtivityb}
\end{subfigure}
\caption{Discrete equilibrium probability distribution of the number of bound particles, $P\left(N\right)$, for $J=-10$ and three different mean occupancies: $\langle\varphi\rangle= 1/3,\, 1/2,$ and $2/3$, respectively. Histograms show the result of exact enumeration based on the 1D Hamiltonian from Eq.~(\ref{eq:Hamiltonian}) for a system of size $L=13$ (a) and $L=14$ (b).}
\label{fig:PDF-strong-anticooperqtivity}
\end{figure}

The situation is more complicated for strong anti-cooperativity ($J \ll -1$) because the system cannot, in general, be reduced to a simple one or two-state system, except near zero and half-filling. 
Furthermore, near HF, the reduction depends on whether $L$ is even or odd (see Fig. \ref{fig:PDF-strong-anticooperqtivity}). 
In the following, we will focus on the left half of the plots in Fig.~\ref{fig:PDF-strong-anticooperqtivity}, i.~e. on occupancies from $N=0$ to $L/2$, the right half being a reflection of  the left through HF owing to the particle-hole symmetry of the system.

For $L$ even, the PDF has a single peak at $N=0$ at zero filling and at $N=L/2$ at half-filling, leading to vanishing standard deviations in these two limiting cases (see Fig. \ref{fig:SigmaPhiVsPhi_L14}). 
At HF, there is only one allowed state, consisting of $L/2$ non-overlapping particle-hole pairs [$(\varphi_i, \varphi_{i+1})  = (1,0)$]  (with a two-fold degeneracy, because the particles can all be on either the even or odd sites). 

For $L$ odd, the PDF still has a single peak at $N=0$ at zero filling, 
but since no single microstate corresponds to HF, there are in this case two peaks with equal (50\%) weight at $N = (L-1)/2$ and $(L+1)/2$, leading to a non-vanishing standard deviation, $\sigma_{-\infty}^{\rm HF} = 1/(2L)$, (see PDF for $L=13$ in Fig. \ref{fig:PDF-strong-anticooperqtivity}). 
For $L$ odd, particle-hole pairs cannot cover the whole system, and defects must appear to fulfill the HF constraint, either an extra hole ($N = (L-1)/2$) or an extra particle ($N = (L+1)/2$). 

\subsubsection{PDF near half-filling: summary}
We can summarize the situation as follows. 
At (or close to) HF, the four different studied cases give rise to very different results for the fluctuations described by the standard deviation and can be considered as clearly distinguishable signatures of different types of particle-particle correlations in small-size systems.
At half-filling, the standard deviation takes on the following values in decreasing order:
(i) for strong cooperativity ($J\to\infty$), $\sigma = 1/2$;
(ii) for no cooperativity ($J=0$), $\sigma = 1/(2\sqrt{L})$; 
(iii) for strong anti-cooperativity ($J\to -\infty$) and $L$ odd, $\sigma = 1/(2L)$; and
(iv)  for strong anti-cooperativity ($J\to -\infty$) and $L$ even, $\sigma = 0$.

In contradistinction to the strong cooperativity limit where the two-state approximation is valid over the whole range of relative occupancy (from zero to full-filling) for both $L$ even and odd, the strong anti-cooperativity limit is more complicated because the one or two-state approximation is only valid for zero and half-filling. 
Although more than two occupancy states are involved between these limits, it is clear that strong anti-cooperativity favors the formation of particle-hole pairs. 
Therefore for even $L$, the system can be approximated by a system of non-overlapping particle-hole pairs.  
For $L$ odd, two extra defect states must be included (an extra particle or extra hole).  
This restricted grand partition function approach, where only the relevant states participating in the strong cooperativity and strong anti-cooperativity limits are retained, is developed in Section~\ref{sec:SM-pdf} of the Methods.

\vspace{3mm}

\subsection{Application to the bacterial flagellar motor (BFM)}\label{sec:data}	
In what follows, we aim to determine whether cooperativity between BFM stator units, mediated  by a (dimensionless) interaction potential $J$, plays a role in their dynamical assembly at the periphery of the rotor.
	
Like most microorganisms, bacteria live in fluid environments with low Reynolds numbers, making them experience a viscous force much larger than the inertial ones \cite{Purcell1977}.
They have subsequently evolved a variety of compatible motility mechanisms that have been widely studied \cite{Kearns2010,Spormann1999,Henrichsen1983,Martinez1999,Dvoriashyna2021,Wadwha2022}.
One type of motility involves rotating one or several flagella that propel bacteria through aqueous media \cite{Nakamura2019,Grognot2021,Brown2012,Gabel2003,Hosking2006,Sowa2014,Ryu2000,Reid2006,Fung1995}.
The BFM is a transmembrane macromolecular complex that consumes the electrochemical potential across the inner bacterial cell membrane to generate torque and set the flagellum in rotary movement \cite{Berg2003,Nirody2017,Kojima2011,Hosking2006,Biquet-Bisquert2021,Fukuoka2009}. 
One of the most compelling properties of the BFM is its ability to change both its conformation and stoichiometry depending on the external medium, allowing it to change the direction of rotation and the magnitude of torque produced \cite{Bai2010, Lele2012,Lele2013, Tipping2013}. 
Its mechanisms of adaptation to external stimuli have been widely studied, becoming a model molecular machine to investigate properties such as mechano-sensitivity \cite{Bai2010,Castillo2013,Nord2017Catch,Lele2013,Hu2017,Wadhwa2021}, chemotaxis \cite{Sourjik2012,Ma2016,Yuan2013} and dynamic subunit exchange \cite{Tusk2018}. 

Torque is generated by inner membrane ion channel complexes called stator units which dynamically bind and unbind to the peptidoglycan (cell wall) at the periphery of the rotor.  
When unbound, they are inactive and passively diffuse in the inner membrane. 
In their bound state, anchored to the peptidoglycan, the ion channels are activated, and, through a mechanism not yet fully understood, apply torque to the rotor \cite{Santiveri2020, Deme2020, Zhu2014,Kojima2009,Tipping2013Quantification,Kojima2018,Terahara2017,Lin2018}.
Precise measurements of the temporal evolution of the angular velocity of the motor of \textit{E. coli}, a direct proxy of the number of bound stator units, have shown that the system can recruit up to $L \approx 13$ stator units and that the system is mechanosensitive in that stator unit stoichiometry scales with the external torque induced by the viscous drag acting on the flagellum turning inside an aqueous medium \cite{Reid2006}.

Much can be learned about the dynamics of the BFM using bead assay measurements. 
In our experiments performed on \textit{E. coli}, we attach a microparticle (bead) to the `hook' (the extracellular portion that joins the motor to the flagellum) via a flagellum stub.
By tracking the bead's off-axis rotation, we can calculate the angular velocity, $\omega$, and the torque produced, $\tau$ (from the relation $\tau = \gamma \omega$, where $\gamma$ is the drag coefficient that increases with the bead's diameter), both of which are a direct proxy for the number of bound stator units.
We can (indirectly) control the mean number of bound stator units at steady state, $\langle N \rangle$, by varying the beads' size and hence the viscous load, because the binding of stators to the BFM is mechanosensitive (see, e.g., \cite{Nord2017Catch}, and references therein). 
We can thus measure the temporal evolution of the number of bound stator units on individual motors, as well as the fluctuations around mean occupancy (for more information on the experimental setup, see references \cite{Nord2017Catch,Perez-Carrasco2022}).

We describe the mechanosensitive binding and unbinding of stator units in the stationary angular velocity regime of the BFM in terms of our adsorption model (see Eq.~(\ref{eq:Hamiltonian})), the rotor being a small-size substrate with periodic boundary conditions onto which up to $L = 13$ stator units can bind at fixed equally spaced positions.
Our working hypothesis, in line with previous studies, is that as long as we only focus on the occupancy of the stator units in the (non-equilibrium) motor stationary state this quantity fluctuates around a fixed mean value as if the stator subsystem were effectively at equilibrium.
In this picture, we account for the presence of the unbound (inactive) stator units diffusing freely in the inner membrane by imposing an external reservoir chemical potential, $\mu_{\rm r}$, which is taken to be constant, as we expect depletion effects to be negligible. 
We incorporate the mechanosensitivity into the model in an average way by assuming that the stator unit binding energy, $\varepsilon$, depends on the viscous load (in our case dependent on the size of the bead and the viscosity of the surrounding medium). 
Load-induced changes in $\varepsilon$ will naturally lead to load-dependent average occupancies and fluctuations,  as observed experimentally.
Furthermore, we assume that the interaction parameter $J$ remains fixed (independent of the load) and check \textit{a posteriori}  if this assumption is consistent with the data.

\begin{figure}
	    \centering
	    \includegraphics[width=130mm]{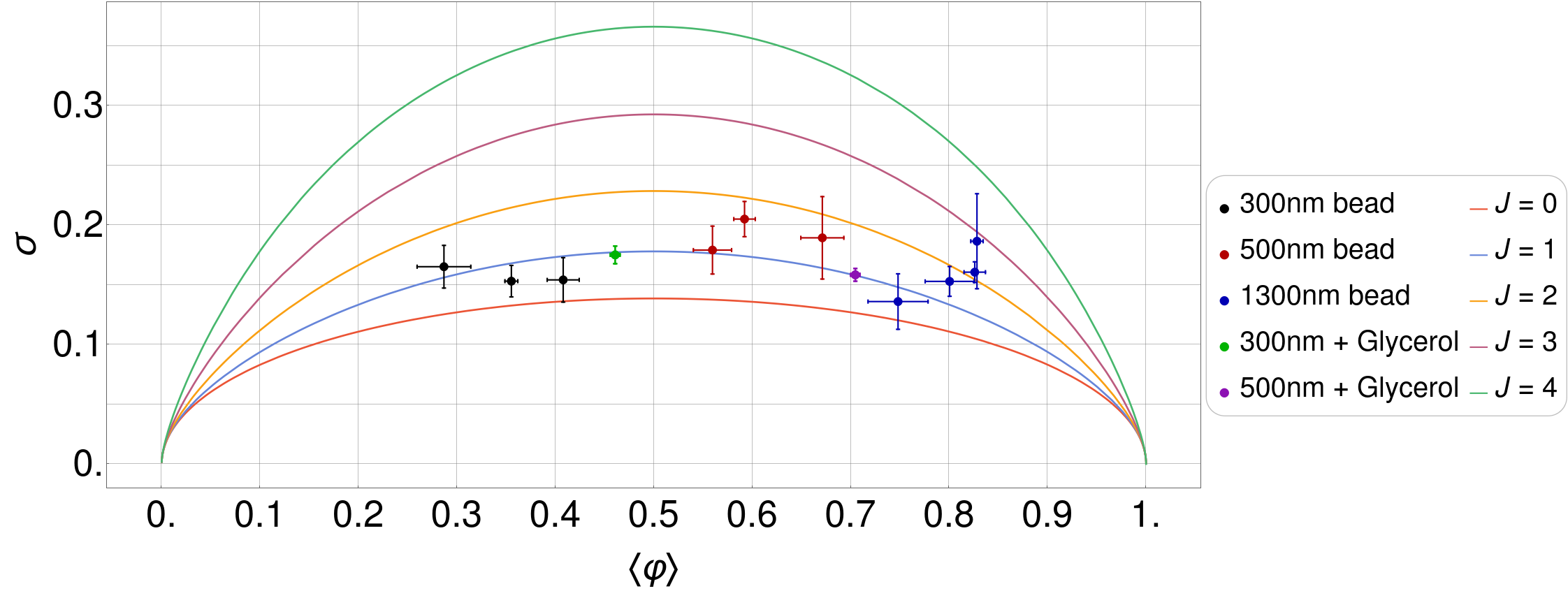}
	    \caption{Parametric plot of the standard deviation of the relative occupancy, $\sigma$, versus the mean occupancy, $\langle \varphi \rangle$, comparing the experimental values with the theory; here, $L=13$. The experimental data corresponds to five different applied viscous loads with three different micro-particles (beads) whose diameter is indicated in the legend.  The experimental data fall between the values of $J = 0.5$ and $J = 2$ of the theoretical curves (aside from one outlier). 
        From this data, we can estimate a characteristic experimental error of $\Delta \sigma \approx 0.02$.}
	    \label{fig:std_vs_phi_EXP}
	\end{figure}

\begin{figure}[ht]
\centering
    \includegraphics[width=86mm]{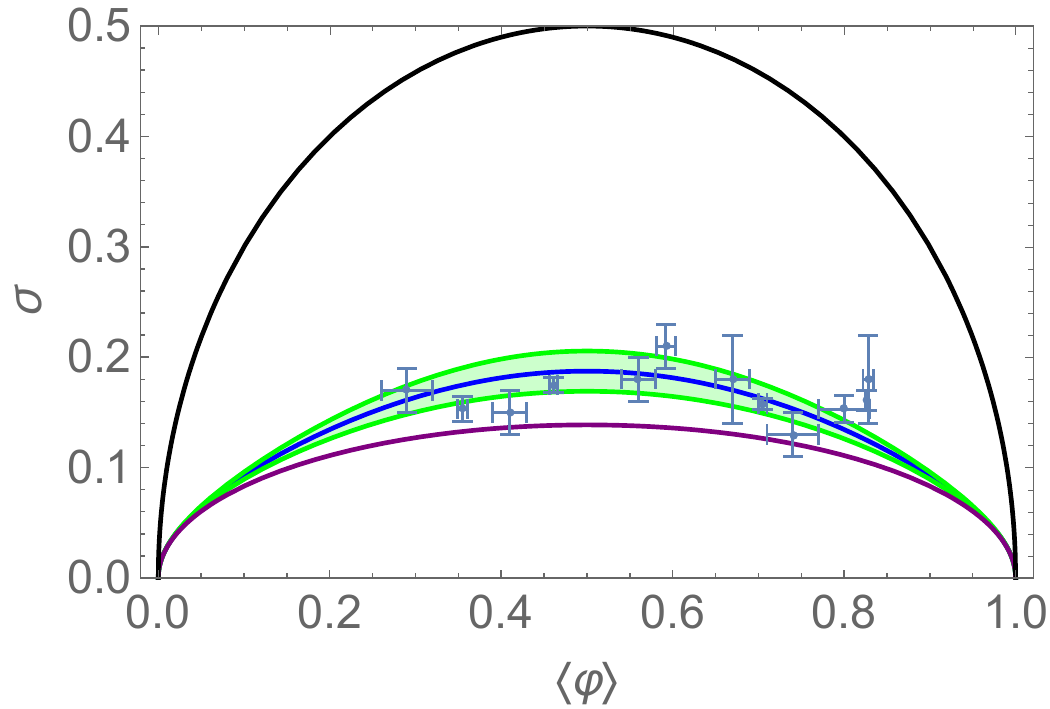}
    \caption{Fit of the theoretical standard deviation (blue solid line) as a function of $\langle \varphi \rangle$ (for $L=13$) to the BFM experimental data (blue data points with experimental error bars). The green shaded area is the  90\% confidence window and the best fit value is  $J = 1.21 \pm 0.22$, where 0.22 is the standard error of the nonlinear fit. The solid purple curve is the result for the Hill-Langmuir model ($J=0$) and the solid black curve is the strong coupling result ($J\to\infty$).}  
    \label{fig:sigmafit}
\end{figure}

As explained previously, we attempt to use the fluctuations in the average number of bound stator units in the stationary
angular velocity regime to determine whether or not cooperativity is at play in the BFM.
Fig. \ref{fig:std_vs_phi_EXP} shows a comparison between the theoretical standard deviation curves, already presented in Fig. \ref{fig:SigmaPhiVsPhi}, and the experimental standard deviations from five different applied viscous loads, corresponding to three different beads with different diameters that are indicated in the legend. 
One can see that the experimental data fall between the values of $J = 0.5$ and $2$, leading to the conclusion that (i) a constant (load-independent) cooperativity parameter ($J$) is a reasonable working hypothesis, (ii) a moderate level of cooperativity in the system, $J \approx 1.21 \pm 0.22$, obtained by fitting the model to the experimental data, is consistent with the experimental observations, see Fig.~\ref{fig:sigmafit} (0.22 is the standard error of the nonlinear fit); and (iii) the estimated value of $J$ is coherent with what is expected for typical biological systems exhibiting moderate cooperativity.

\begin{figure}
\centering
\begin{subfigure}{0.32\textwidth}
\includegraphics[width=\textwidth]{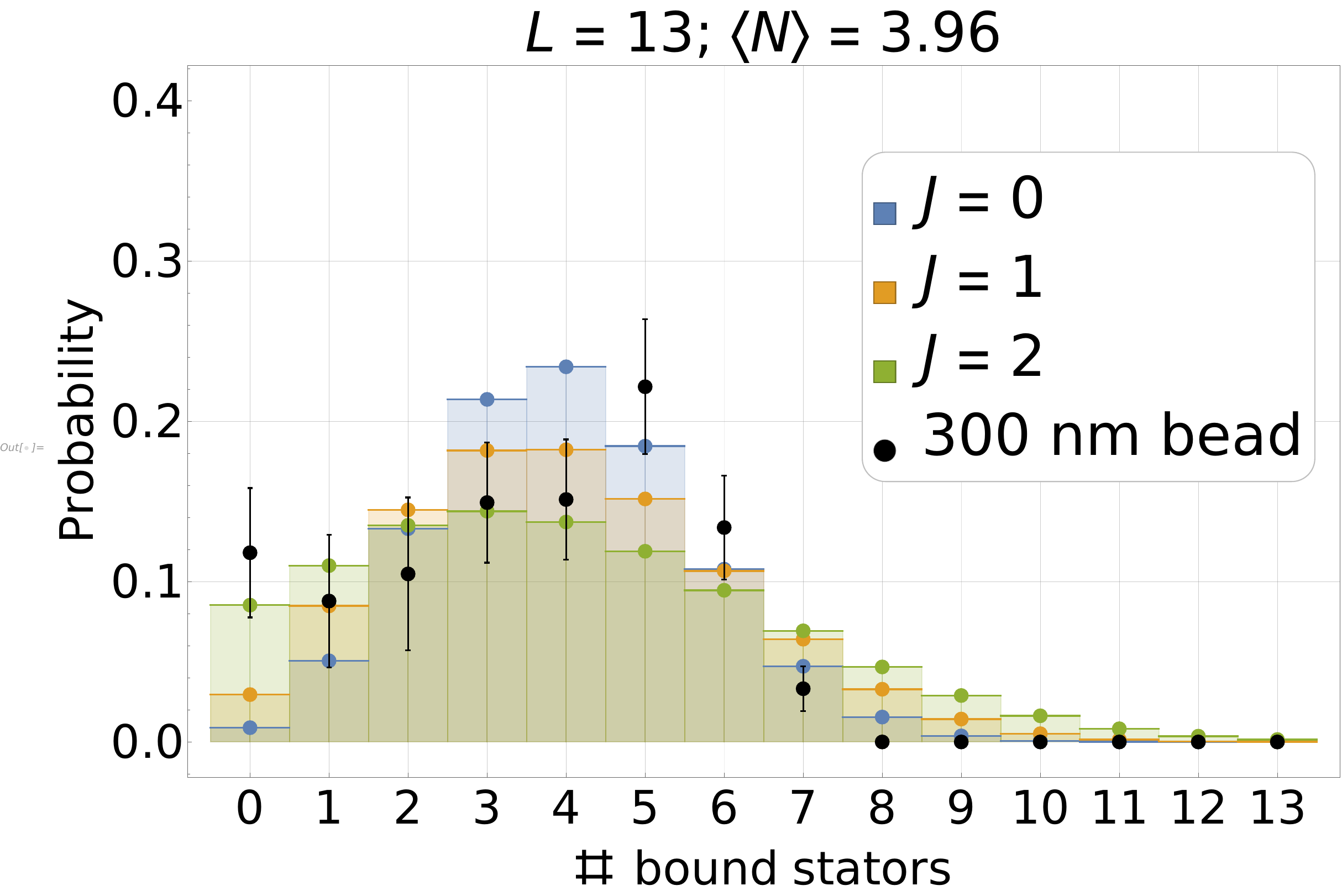}
\caption{}\label{fig:OccupancyPDF_EXPa}
\end{subfigure}
\begin{subfigure}{0.32\textwidth}
\includegraphics[width=\textwidth]{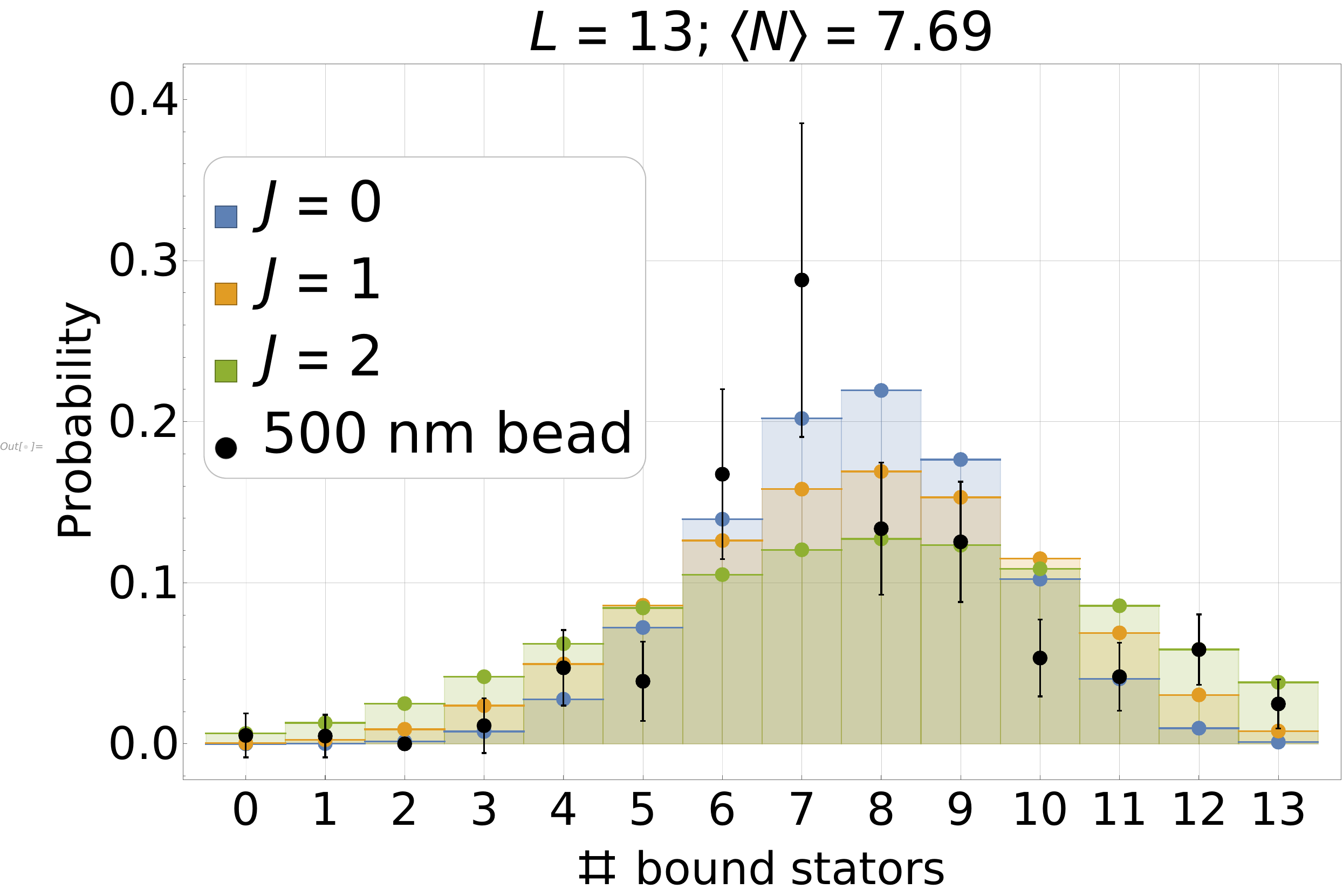}
\caption{}\label{fig:OccupancyPDF_EXPb}
\end{subfigure}
\begin{subfigure}{0.32\textwidth}
\includegraphics[width=\textwidth]{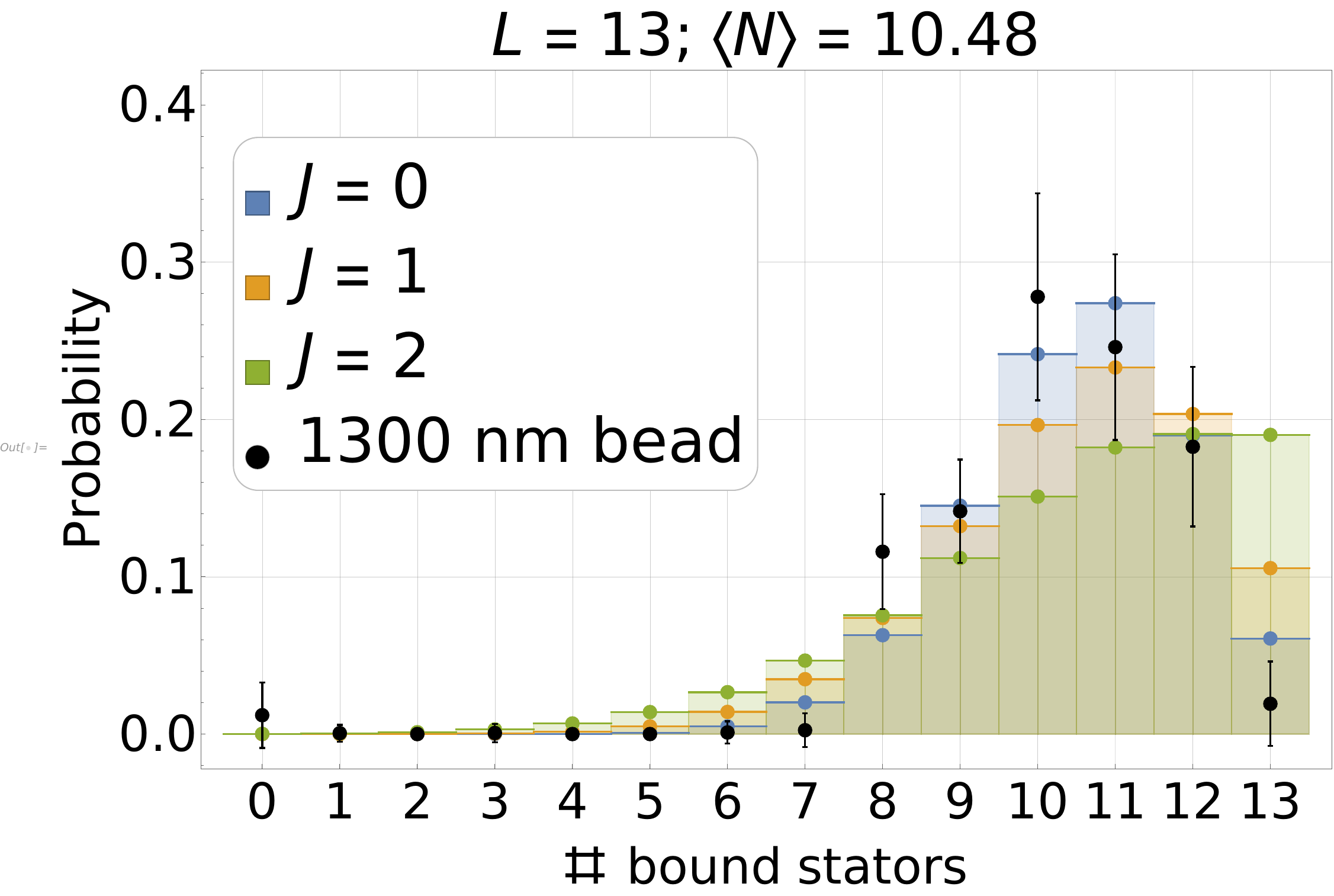}
\caption{}\label{fig:OccupancyPDF_EXPc}
\end{subfigure}
\caption{Discrete equilibrium probability distribution of bound stator units, $P(N)$, for the experiments (red with black dots) and different values of the interaction potential, $J = 0$ (blue), $1$ (orange), $2$ (green), obtained from exact enumeration of states from the effective Hamiltonian; here, $L=13$. The mean equilibrium occupancy of each distribution is indicated on top of the three plots, which corresponds to a given external load imposed by beads of different diameters indicated in the legend.}
\label{fig:OccupancyPDF_EXP}
\end{figure}
 
To delve into further detail, we can also compare the probability distributions of the occupancy shown in Section~\ref{sec:ProbDist} with our preliminary experimental ones. Fig. \ref{fig:OccupancyPDF_EXP} shows a comparison between the probability distribution from experiments (black dots) and the probability distributions obtained by exact enumeration for different values of $J$.
Even though, in this analysis of preliminary data, we don't observe a precise fit with any of the chosen values, it is clear that the PDFs for the range of $J$ estimated from the standard deviation are coherent with our experimental results. 
This result suggests that the interaction potential has a non-zero positive value qualitatively in accordance with the value range deduced from the above fluctuation analysis.
Moreover, as further explained in the Methods section, we observe theoretically that the system becomes bimodal (with only the completely full and empty states having significant probabilities) for slightly higher values of the interaction potential, $J \gtrsim 5$, behavior that is not seen in the experimental PDF data. We see from Section \ref{sec:ProbDist} (Fig.~\ref{fig:OccupanyPDF}) that for the chosen system size ($L=13$), $J\approx3$ is a threshold value marking a transition from low cooperativity PDFs having a maximum centered on the average occupancy to high cooperativity PDFs exhibiting a bimodal form with local maxima at zero and full filling. In the latter case, microstates with fillings close to the average occupancy have low probabilities, and the motor would undergo discontinuous jerky motion. Such a bacteria would fluctuate between an immobile state (zero stator unit occupancy) and maximum speed (full stator unit occupancy).
Knowing that the BFM is a highly and smoothly adaptive molecular machine, such high values of cooperativity would, therefore, not be expected since such a motor would not be able to adapt smoothly to environmental variations. 

\section{Discussion and Conclusion}
We have focused on cooperative processes involving the adsorption of ligands onto a substrate disposing of a limited number of binding sites.
By presenting a general method (based on a 1D periodic lattice gas) for recognizing and assessing characteristic signatures of cooperativity or anti-cooperativity in the stochastic occupancy fluctuations, we propose both a criterion to determine whether any given adsorption system exhibits cooperative or anti-cooperative behavior and a method to quantify the amplitude of the ligand-ligand interaction potential.

In the process, knowing that in the thermodynamic limit relative occupancy fluctuations (or standard deviation) vanish, we have addressed the following essential questions  :
(i) what is a sufficiently “small” system for studying fluctuations?
(ii) what model parameter values allow a system to smoothly “adapt” to external conditions?

We compared the theoretical results for both the standard deviation and the probability distribution function of stator unit occupancy with experimental data obtained for the BFM.
We concluded that a moderate value of cooperativity, $\approx 1 \,k_\text{B}T$  (i.e., $J \approx 1$ with a standard error of about $20\%$), for the short-range (SR) model is not only coherent with the experimental data but also expected from the characteristic smooth adaptability of the motor to changing external loads [for the infinite-range (IR) model, using the mapping (\ref{EQ:mapping}) derived in Section~\ref{sec:IR-model}, the corresponding value is $J_{\rm IR} = 2 J_{\rm SR}/(L-1) \approx 0.20$]. 
For the characteristic size of the BFM ($\approx 10$ binding sites), slightly higher values of cooperativity would lead to a motor that is bimodal in the occupancy and, therefore, would exhibit switch-like behavior for the produced torque not compatible with the required motor characteristics. 
The highly cooperative bimodal behavior found at strong coupling does, however, appear to describe the rapid stochastic switching in BFM rotational direction \cite{Duke2001ConformationalSpread,Bai2010,Bai2012Coupling}.

Within the framework that we have developed, we are now in a position to propose, using the BFM as an illustrative example, a general principle of motor \textit{adaptability} depending on whether the motor under investigation should respond smoothly to external stimuli or behave like a two-state switch. 
As stated earlier, we suppose that for the BFM the binding energy $\varepsilon$ depends on the load (and therefore the bead size). A change in load would therefore lead directly to a change in the effective chemical potential $\mu$ and therefore a modification of the average occupancy and standard deviation (fluctuations). 

\begin{figure}[ht]
    \includegraphics[width=86mm]{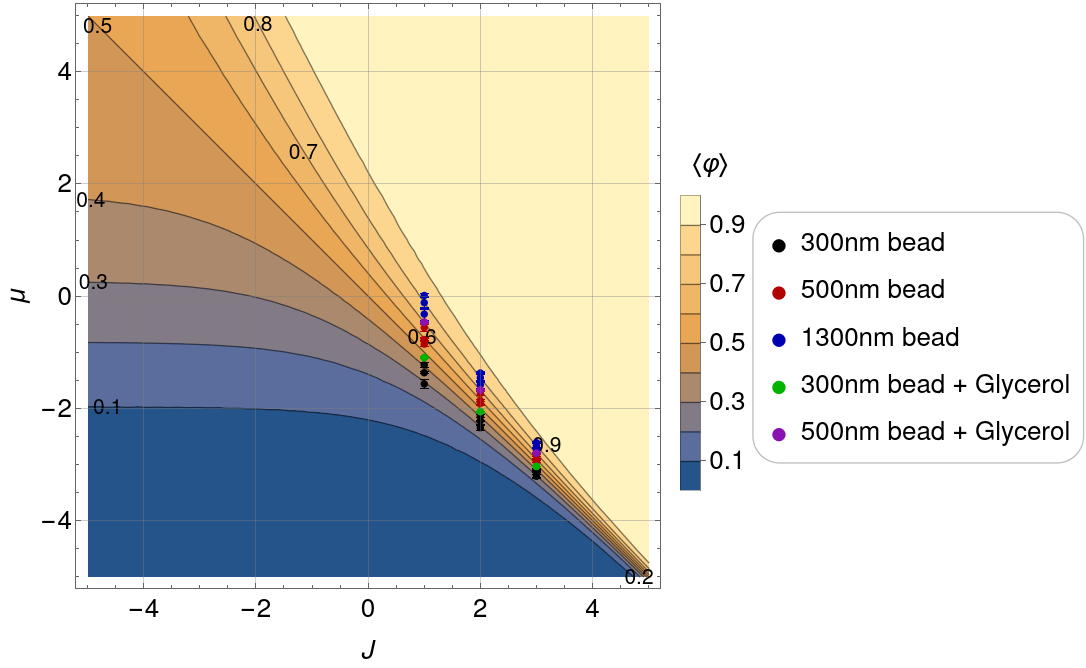}
    \includegraphics[width=86mm]{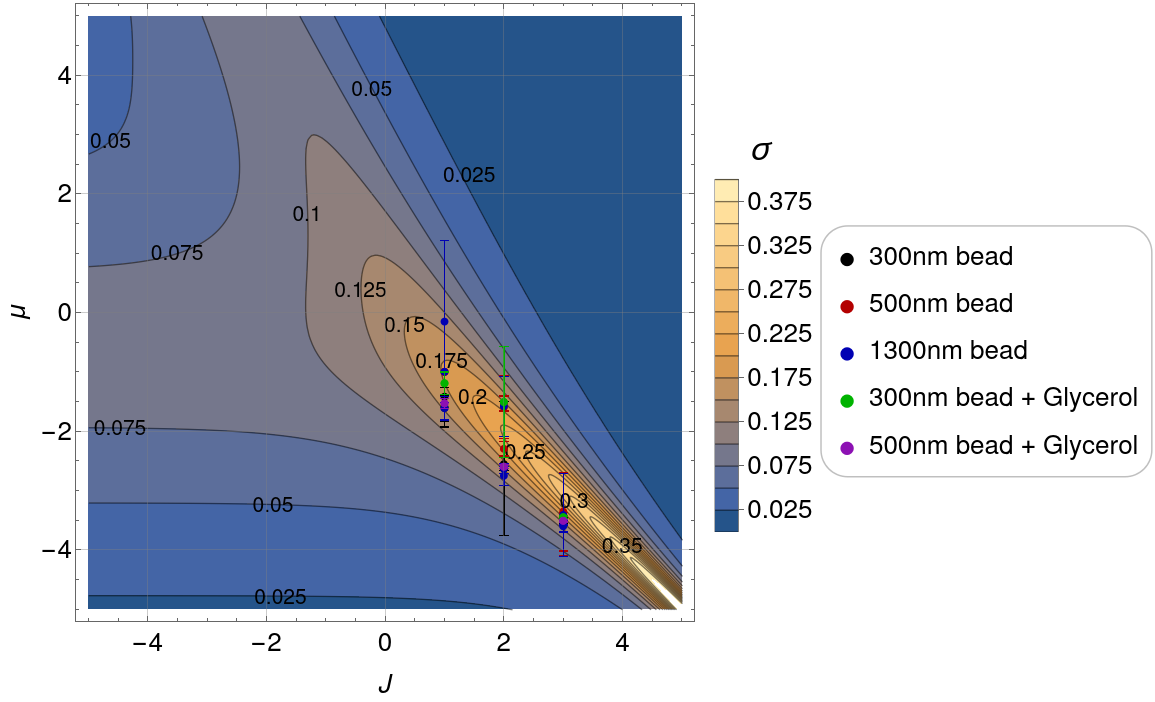}
    \caption{Values of the effective chemical potential $\mu$ extracted from the experimental data for different values of the interaction potential $J$ ($L=13$). We situate them on  the contour plots of $\langle \varphi \rangle$ and $\sigma$ in order to find the interval of chemical potential for which the BFM would work at fixed $J$.}
    \label{fig:Phi+SigmaPhi+data}
\end{figure}

For the short-range model, these modifications to $\mu$ (extracted from the experimental data) are shown in Fig.~\ref{fig:Phi+SigmaPhi+data} on the occupancy and standard deviation contour plots for each of the three studied values of $J$  (for $L=13$).  
By positioning the effective chemical potential $\mu$ window in this way we observe that for $1 <  J < 2$ it is situated in a \textit{sweet spot} suitable for a motor that responds smoothly to environmental changes with the ability to cover a wide range of occupancies spanning half-filling while minimizing the amplitude of the fluctuations. 

For the BFM the occupancy is directly related to the motor speed and therefore, near this \textit{sweet spot}, the BFM can smoothly adjust its speed in response to external stimuli with a minimum of fluctuations. In retrospect the bead sizes used in the experiments were clearly chosen to \textit{see} an effect because a much higher load would have forced the system into nearly full filling (exactly the case in the stall experiments reported in \cite{Perez-Carrasco2022});  a much lighter load would have pushed the system to nearly zero filling (similar to what was done using another technique in the resurrection experiments).
The highly cooperative switch-like behavior observed for the BFM rotational direction \cite{Duke2001ConformationalSpread,Bai2010} would place this system in the strong coupling regime on the right-hand side of the contour plots ($J>4$).

For the motor to respond smoothly to external stimuli \textit{and} cover a wide range of occupancy with minimum of fluctuations we therefore see that moderate positive values of cooperativity $1 <  J < 2$ are optimal given the $\mu$ window imposed by the system characteristics.
Although the occupancy range would increase at higher values of $J$, this positive effect would be counterbalanced by a strong increase in fluctuations because the system would be pushed into the switch-like operation regime. 
On the other hand, for lower and even negative values of $J$, the fluctuations would (favorably) be diminished in amplitude, but at the cost of severely restricting the accessible range of occupancy (and therefore motor speed for the BFM). 

By examining Fig.~\ref{fig:Phi+SigmaPhi+data}, we observe that if a two-state switch-like motor operation were sought after in order to cover a wider range of average relative occupancy, then higher values of $J$ would be the best choice, the price to pay would be a strong increase in fluctuations, provided that the observation time window be wide enough.  
If a motor with weak fluctuations and a restricted range of occupancies were sought after then negative values of $J$ (anti-cooperativity) would be the best choice. In fact, this type of motor would be relatively immune to changes in external stimuli and therefore exhibit a relatively constant speed (provided the speed were still directly related to occupancy, as for the BFM). 

We have not addressed here the biochemical origins of stator-stator interactions, although one can imagine that stator units interact at short range much like proteins, either directly or through allosteric pathways, and that long-range interactions could also be due to allosteric effects. 
We plan to address this and other open questions in future work. 
In the present context the major important open question in modeling the BFM concerns how to integrate cooperativity into kinetic models that already allow one to account for the relaxation time asymmetry between stall and resurrection \cite{Perez-Carrasco2022}. The goal is to arrive at a model that can describe in a unified way both BFM stator number fluctuations and relaxation time asymmetries.

\section{Methods}\label{sec:methods}
\subsection{1D short-range lattice gas (SRLG) \label{sec:SRLG}}	
Analytical solutions for the equilibrium state can be obtained using the transfer matrix formalism \cite{kardar2007,mccoy2014two}. According to it, the grand partition function of the system can be written as:
	\begin{equation}    
    \Xi = \sum_{\{\varphi_i\}} \exp [-\beta \mathcal{H}(\bm{\varphi})]
    = \sum_{\{\varphi_i\}} \prod_{i=1}^L \bra{\varphi_i}T\ket{\varphi_{i+1}},
    \label{PartFucTM}
\end{equation}
where $T$ is the transfer matrix, which for the system we are working on, is given by
\begin{equation}
    T = 
    \begin{pmatrix}{}
        1 & \e^{\frac{\mu}{2}} \\
        \e^{\frac{\mu}{2}} & \e^{(J + \mu)}
    \end{pmatrix}.
    \label{TM}
\end{equation}
In the case of periodic boundary conditions, equation \eqref{PartFucTM} can be simplified to 
\begin{equation}
\Xi = \text{Tr} \ T^L = \lambda_+^L + \lambda_-^L  
\label{eq:PartFunction}
\end{equation}
where 
	\begin{equation}
	\lambda_{\pm} = \e^{X} \left(\cosh{X} \pm \sqrt{\sinh^2{X} + \e^{-J}} \right)
	\label{eigenvalues}
	\end{equation}
are the eigenvalues of the transfer matrix and $X = \frac{1}{2}(J + \mu)$. 
From this result \eqref{eq:PartFunction} for the grand partition function, one can obtain the mean relative occupancy, $\langle\varphi\rangle$, and the mean square relative occupancy, $\langle\varphi^2\rangle$,
as functions of $\mu$ and $J$
by employing known relations between thermodynamic quantities 
(i.e., by taking derivatives of $\Xi$ with respect to the effective chemical potential $\mu$). The standard deviation $\sigma$ can then be derived 
as a function of $\mu$ and $J$ from these two averages. 

The transfer matrix formalism can also be used to calculate the occupancy-occupancy correlation function, which provides another 
route to $\langle\varphi^2\rangle$ (and therefore
$\sigma$), 
but now as an explicit function of $\langle\varphi\rangle$ and the correlation length, 
$\xi \equiv 1/\ln\left(\lambda_+/\lambda_-\right)$ (already presented in Eq.~\ref{eq:xi}).
This second form for $\sigma$ lends itself to a clearer physical picture of how the standard deviation interpolates between the zero and infinite cooperativity limits.

According to the grand canonical formalism, the mean relative occupancy of a system is given by the expression
	\begin{equation}
	    \langle \varphi \rangle 
     =\frac{1}{L}\frac{1}{\Xi}\pa{\Xi}{\mu}{}
     = -\frac{1}{L^2}\pa{\Omega}{\mu}{},
     \label{eq:Dphi}
	\end{equation}
	where $\Omega = -L\ln{\Xi}$ is the grand potential corresponding to the grand partition function. 
 Using the chain rule, we can find that
	\begin{equation}
	    \langle \varphi \rangle = \frac{1}{\Xi} \left(\lambda_+^{(L-1)}\pa{\lambda_+}{X}{}\pa{X}{\mu}{} + \lambda_-^{(L-1)}\pa{\lambda_-}{X}{}\pa{X}{\mu}{} \right).
	\end{equation}
 Replacing 
 \begin{equation}\pa{X}{\mu}{} = \frac{1}{2} \quad \text{and}\quad  \pa{\lambda_\pm}{X}{} = \lambda_\pm\left(1 \pm \frac{\sinh{X}}{\sqrt{\sinh^2{X} + \e^{-J}}}\right)
 \end{equation}
in the previous equation, we obtain Eq.~\ref{eq:phi}.
	
Similarly, we can determine the mean square relative occupancy by taking another derivative:
	\begin{equation}
	    \langle \varphi^2 \rangle = \frac{1}{L^2}\frac{1}{\Xi}\pa{\Xi}{\mu}{2}.
     \label{eq:Dphi2}
	\end{equation}
By carrying out this calculation, we obtain:
	\begin{equation}
	    \langle \varphi^2 \rangle = \frac{1}{4}\Biggl[1+\frac{\sinh^2{{X}}}{\sinh^2{{X}}+\e^{-J}} 
     + \tanh{\left(\frac{L}{2{\xi}} \right)} \left(\frac{2 \sinh{{X}}}{\sqrt{\sinh^2{{X}}+\e^{-J}}} \right. + \left.  \frac{1}{L}\frac{\e^{-J} \cosh{{X}}}{(\sinh^2{{X}} + \e^{-J})^{3/2}}\right) \Biggl] \label{phi2}.
	\end{equation}

Denoting $\xi$ as the correlation length makes physical sense because correlations decay exponentially on a scale $\xi$, as can be seen by inspecting the correlation function obtained using the transfer matrix formalism:
\begin{equation}
\langle \varphi_i\varphi_{i+r} \rangle = \frac{1}{K}\left[\lambda_+^L\left(1 + \lambda_+^{-r}\lambda_-^r v_-^2\right) + \lambda_-^L v_-^2\left(v_-^2 + \lambda_+^r\lambda_-^{-r}\right)\right],
    \label{eq:twopointcorr}
\end{equation}
where $v_- = \e^{\mu/2}(\lambda_- - 1)$ and $K =(\lambda_+^L + \lambda_-^L)(1 + v_-^2)^2$. 
By using the definition of $\xi$ presented above, the products of $\lambda_+^r \lambda_-^{-r}$ and $\lambda_+^{-r}\lambda_-^r$ can be rewritten as $\e^{\pm r/\xi}$.
This result for the correlation function, Eq.~\ref{eq:twopointcorr}, can be put into a more physically transparent form, 
\begin{equation}
\langle \varphi_i\varphi_{i+r} \rangle = \langle \varphi \rangle +
\frac{\langle \varphi \rangle_{\rm TL} - \langle \varphi \rangle_{\rm TL}^2}
{1+\e^{-L/\xi}} 
\left[
\e^{-r/\xi} + \e^{-(L-r)/\xi} - 1 - \e^{-L/\xi}
\right].
 \label{eq:corrfct}
\end{equation}
This result, which illustrates how the correlation function can be written explicitly in terms of $\langle  \varphi \rangle$, $\xi$, and $L$, allows the exponential decay on a scale $\xi$ to be made clear.
Here $\langle \varphi \rangle_{\rm TL}$ is 
the average relative occupancy in the thermodynamic limit (TL),
defined to be an 
explicit function of $\langle  \varphi \rangle$ and $\xi$:
\begin{equation}
\langle \varphi \rangle_{\rm TL} 
\equiv
\frac{1}{2} 
\left[1 + \coth \left( \frac{L}{2\xi} \right)  (2 \langle \varphi \rangle - 1) \right].
\label{eq:phiTL}
\end{equation}
This result for the correlation function, Eq.~(\ref{eq:corrfct}), is clearly periodic and tends to the correct TL when  $L\to\infty$:
\begin{equation}
\langle \varphi_i\varphi_{i+r} \rangle_{\rm TL} = 
\langle \varphi \rangle_{\rm TL}^2 +
\left(\langle \varphi \rangle_{\rm TL} - \langle \varphi \rangle_{\rm TL}^2\right)
\e^{-r/\xi}.
 \label{eq:corrfcttl}
\end{equation}

By performing the sum depicted in Eq.~\ref{eq:phi2sum} using Eq.~\ref{eq:twopointcorr} 
we obtain for the mean square relative occupancy an expression that explicitly depends on the eigenvalues and the correlation length as follows:
\begin{equation}
    \langle \varphi^2 \rangle = \frac{1}{K}\left[\lambda_+^L \left(1 + \frac{v_-^2}{L}\frac{1 - \e^{-L/\xi}}{1 - \e^{-1/\xi}}\right) + \lambda_-^L v_-^2\left(v_-^2 + \frac{1}{L} \frac{1 - \e^{L/\xi}}{1 - \e^{1/\xi}} \right) \right].
    \label{eq:sumcorr}
\end{equation} 
It can be shown that Eq.~\ref{eq:sumcorr} is equivalent to Eq.~\ref{phi2}, which provides a check on the thermodynamic result.  
In the calculation leading to Eq.~\ref{phi2}, the correlation length appears from the fraction $(\lambda_+^L - \lambda_-^L)/(\lambda_+^L + \lambda_-^L)$ which can be rewritten as $\tanh(L/(2\xi))$.  

With both Eq.~\ref{eq:phi} and Eq.~\ref{phi2}, we can obtain an analytical expression for the standard deviation from the expression $\sigma = \sqrt{\langle \varphi^2 \rangle - \langle \varphi \rangle^2}$.  
The result obtained in this way is equivalent to the one obtained more efficiently in the main text using the thermodynamic relation
$\sigma = L^{-1/2}\sqrt{\partial_\mu \langle \varphi \rangle}$.

\begin{figure}
    \centering
    \includegraphics[width=86mm]{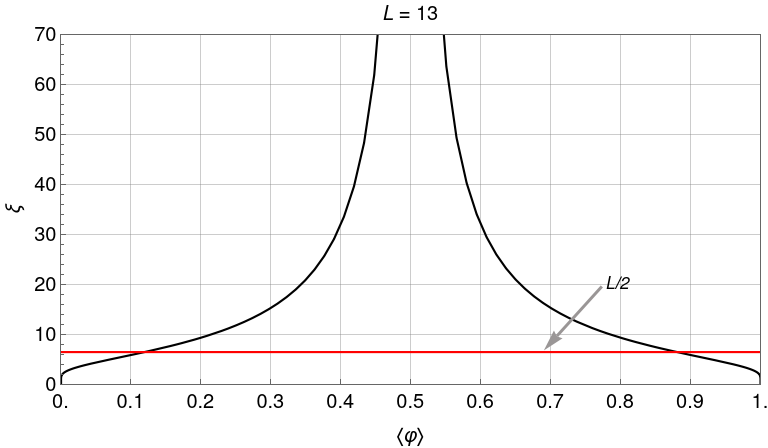}
    \caption{Correlation length, $\xi$, in the limit of $J\rightarrow \infty$. It diverges at half-filling and is mainly higher than $L/2$ (indicated by the red line), except near zero and full fillings.}
    \label{fig:xiJinf}
\end{figure}

To make explicit the $\xi$ dependence we now calculate $\sigma$ using the  \textit{connected} correlation function, given by
$\langle \varphi_i \varphi_j \rangle  - \langle \varphi  \rangle^2$,
in the double sum leading to $\sigma^2$:
\bea
\sigma^2 & = &
\langle\varphi^2\rangle - \langle\varphi\rangle^2 =
\frac{1}{L^2} \sum_{i=1}^{L} \sum_{j=1}^{L} 
\left[
\langle \varphi_i \varphi_j \rangle  - \langle \varphi  \rangle^2
\right] \\
& = & \frac{\langle \varphi \rangle}{L}-\langle \varphi \rangle^2 +
\frac{2}{L}  \sum_{r=1}^{L-1} \left(1 - \frac{r}{L}\right) 
\langle\varphi_1 \varphi_{1+r}\rangle.
\label{eq:sigmacorrfct}
\eea
The sum in \ref{eq:sigmacorrfct} can be evaluated using 
Eq.~\ref{eq:corrfct}.
to obtain a physically transparent expression for the variance, $\sigma^2$, explicitly in terms of $\langle \varphi \rangle$, $\xi$, and $L$:
\begin{equation}
\sigma^2 = \sigma_\infty^2
+ \left(
\langle \varphi \rangle_{\rm TL} - \langle \varphi \rangle_{\rm TL}^2
\right)
F(\xi;L),
\label{eq:sigmacorr}
\end{equation}
where
$\langle \varphi \rangle_{\rm TL}$ is given by Eq.~\ref{eq:phiTL}
and
\begin{equation}
F(\xi;L) \equiv 
\frac
{
(L+1) \left(  \e^{-\frac{1}{\xi}} - \e^{-\frac{L}{\xi}} \right)
+
(L-1) \left(  \e^{-\frac{(L+1)}{\xi}} - 1 \right)
}
{
L \left(
1-\e^{-\frac{1}{\xi}}
\right)
\left(
1+\e^{-\frac{L}{\xi}}
\right)
}.
\end{equation}
This form for the variance allows the weak and strong cooperativity limits to be clearly identified, since $F \to 0$, when $J\to \infty$ and $F \to -1 + 1/L$, when $J\to 0$ 
(in the latter limit, $\langle \varphi \rangle_{\rm TL} \to \langle \varphi \rangle$).
More generally, for fixed $J$, $\xi$ becomes an implicit function of  $\langle \varphi \rangle$ and can be plotted parametrically, as in Fig.~\ref{fig:xiJinf} for infinite $J$. 
 
The  thermodynamic limit corresponds to $L \rightarrow \infty$, for which $\tanh{(L/2\xi)} \rightarrow 1$, therefore the average relative occupancy simplifies to
	\begin{equation}
	\langle \varphi \rangle_{\rm TL} = \frac{1}{2}\left(1 + \frac{\sinh{X}}{\sqrt{\sinh^2{X} + \e^{-J}}} \right), \label{averageTL}
	\end{equation}
and $\langle \varphi^2 \rangle_{\rm TL} = \langle \varphi \rangle^2_{\rm TL}$. Hence the standard deviation in the thermodynamic limit, $\sigma_{\rm TL}$, is zero and self-averaging holds.

\subsection{Probability distribution function of the occupancy}
\label{sec:SM-pdf}
 For high positive values of $J$, e.~g. $J\geq5$, the PDF saturates by accumulating at the boundaries to 
\be
P_\infty\left(N; \langle\varphi\rangle \right) =  (1-\langle\varphi\rangle) \delta_{N,0}
+  \langle\varphi\rangle \delta_{N,L}
\ee
and the system becomes well described by an effective two-state system.  
The standard deviation for the occupancy $N$ therefore saturates  at $L\sigma_\infty$  ($=L/2$ at half-filling), where we recall that $\sigma_\infty = \sqrt{\lan\varphi\ran - \lan\varphi\ran^2}$, and it becomes  easy to calculate all moments of $N$:   
$\langle N^m \rangle = L^m \langle \varphi \rangle$ 
(or $\langle \varphi ^m \rangle = \langle \varphi \rangle$).
The moments are defined by
\be
\mu_n = \lan (\varphi - \lan\varphi\ran)^n \ran
\ee 
and the standard moments by $\mu_n/\sigma^n$, where $\sigma = \sqrt{\mu_2}$ is the standard deviation.
In the strong cooperativity limit, one can therefore find simple explicit  expressions for low order standard moments, such as the skewness $\gamma = \mu_{3}/\sigma^3$,
\be
\gamma_\infty = \frac{\mu_{3,\infty}}{\sigma_{\infty}^3} = 
\frac{\lan\varphi\ran - 3 \lan\varphi\ran^2 + 2 \lan\varphi\ran^3}{\sigma_\infty^3}  
\ee
and kurtosis, $\kappa = \mu_{4}/\sigma^4$,
\be
\kappa_{\infty} = \frac{\mu_{4,\infty}}{\sigma_\infty^4} = 
\frac{\lan\varphi\ran - 4 \lan\varphi\ran^2 + 6 \lan\varphi\ran^3- 3 \lan\varphi\ran^4}{\sigma_\infty^4}.
\ee
These results for the effective two-state system are very different from those predicted for a Gaussian PDF.

For strong anti-cooperativity ($J \ll -1$),  the system near zero and half-filling can be reduced to a simple one or two-state system. Near HF, the reduction depends on whether $L$ is even or odd (see Fig. \ref{fig:PDF-strong-anticooperqtivity}). The PDF always has a single peak at $N=0$ at zero filling.
For even $L$ in the strong anti-cooperativity limit, the PDF also has a single peak at $N=L/2$  at half-filling, 
\be
P_{-\infty} (N; 1/2) =  \delta_{N,L/2}    \qquad (L \, \rm{even}),
\ee
(see Fig. \ref{fig:PDF-strong-anticooperqtivity})
leading to a vanishing standard deviation.
For $L$ odd at half-filling, on the other hand, there are two peaks with equal (50\%) weight at $N = (L-1)/2$ and $(L+1)/2$, 
\be
P_{-\infty} (N; 1/2) =  \frac{1}{2}\delta_{N,(L-1)/2} + \frac{1}{2}\delta_{N,(L+1)/2} \qquad  (L \, \rm{odd}),
\ee
leading to a non-vanishing standard deviation of $1/2$ for the occupancy $N$.
Unlike for $L$ even, for $L$ odd non-overlapping particle-hole pairs cannot cover the whole system, and a defect \emph{(non-particle-hole pair)} must appear, either an extra hole [$N = (L-1)/2$] 
or an extra particle [$N = (L+1)/2$], to arrive at HF.

The restricted grand partition function approach consists in keeping in the sum over all possible microstates $\{ \varphi_i \}$ only those states that survive in the studied limit.
Before presenting this approach, we first recall the exact calculation in the absence of cooperativity ($J=0$). 

For $J=0$
the partition sum can be organized into a sum over states with a fixed  occupancy $N$ 
with the Boltzmann factor 
$\exp [-\beta \mathcal{H}_0(N)]=\e^{\mu N}$
multiplied by a multiplicity (or binomial coefficient)  
 $C^L_N = 
 L! /\left[N! \left(L-N\right)!\right]$ (related to the configurational entropy) that gives the number of microstates $\bm{\varphi}$ consistent with occupancy $N$:
\be
\label{eq:zo}
\Xi_0 = \sum_{\{ \varphi_i \}} 
\e^{ -\beta \mathcal{H}_0 \{ \varphi_i \} } =  \sum_{N=0}^L C^L_N \e^{\mu N} = 
\left( 1 + \e^\mu  \right)^L.
\ee
From $\Xi_0$ we immediately obtain the expected $J=0$ results:
\be
\langle\varphi\rangle_0 = (L \Xi_0)^{-1}\partial_\mu \Xi_0 = 1/\left( 1 + \e^{-\mu}  \right)
\ee
and
 \be
 \sigma_0 =  \sqrt{L^{-1} \partial_\mu \langle \varphi \rangle_0} = 
 L^{-1/2}  \sqrt{\langle \varphi \rangle - \langle \varphi \rangle^2}.
 \ee
 By introducing the usual (configurational) entropy of mixing, 
 \be
 S_{\rm{mix}} = 
 -k_B L [\varphi \ln (\varphi) + (1-\varphi) \ln (1-\varphi) ],
 \ee
 the multiplicity $C^L_N$ can be accurately approximated (using Stirling's formula) for $L \gg 1$ and $0 < \varphi = N/L < 1$ by
 \be
 \label{eq:multapp}
 C^L_N \approx  \frac{\e^{L s_{\rm{mix}}(\varphi)}}{\sqrt{2 \pi L \varphi (1-\varphi)}},
 \ee
 where 
 \be
 s_{\rm{mix}} \equiv S_{\rm{mix}}/(L k_B) =
 -[\varphi \ln (\varphi) + (1-\varphi) \ln (1-\varphi)]
 \ee
 is the dimensionless entropy of mixing. The entropy of mixing is a non-monotonic (concave) function of $\varphi$ that goes to 0 for zero and full filling and reaches a maximum at HF.
 
 In Eq.~\eqref{eq:multapp} for $L \gg 1$, the main variation comes from the exponential, and it is possible under certain conditions to replace $\varphi$ in the prefactor by $\lan \varphi \ran$ without loss of accuracy.
 The PDF for $J=0$ can then be accurately approximated  by
  \be
 P_0(N; \lan \varphi \ran) \equiv \frac{C^L_N \e^{\mu N}}{\Xi_0} 
 \approx  
 \frac{ \e^{-\beta \mathcal{F}_0 (\varphi;  \lan \varphi \ran)}}
 {(2 \pi)^{1/2} \sigma_0 L \Xi_0(\lan \varphi \ran)},
 \ee
where 
\be
\beta \mathcal{F}_0 (\varphi;  \lan \varphi \ran) =
-L \left[ \varphi \ln \left( \frac{\lan \varphi \ran}{1-\lan \varphi \ran} \right)  +  
s_{\rm{mix}}(\varphi) \right]
\ee
is an effective free energy for $\varphi$
and we have written
\be
\Xi_0(\lan \varphi \ran) = 
\frac{1}{( 1-\lan \varphi \ran)^L},
\ee
and 
\be
\mu(\lan \varphi \ran) = \ln \left( \frac{\lan \varphi \ran}{1-\lan \varphi \ran} \right)
\ee
as functions of $\lan \varphi \ran$.

For $L \gg 1$, we can use the saddle point approximation to obtain a simple Gaussian form for $P_0(N; \lan \varphi \ran)$: by expanding 
$\beta \mathcal{F}_0 (\varphi;  \lan \varphi \ran)$ around $\varphi = \lan \varphi \ran$ (the position of its minimum) to second order in $(\varphi - \lan \varphi \ran)$ we find
\be
 P_0(N; \lan \varphi \ran) \approx  
 \frac{\exp \left[ -\frac{(\varphi - \lan \varphi \ran)^2}{2 \sigma_0^2} \right]}
 {{(2 \pi)^{1/2} \sigma_0 L}},
 \ee
where the variance
\be
\sigma_0^2 = - [L s''_{\rm{mix}}(\lan \varphi \ran)]^{-1} = (\lan \varphi \ran - \lan \varphi \ran^2)/L
\ee
is related to the second derivative of the entropy of mixing, or curvature, evaluated at $\lan \varphi \ran$.
Since $-s''_{\rm{mix}}(\lan \varphi \ran)$ has a minimum at half-filling, $\sigma_0$ has a maximum there. 

The Gaussian approximation for $P_0(N; \lan \varphi \ran)$  is (approximately) correctly  normalized because the sum of the exact PDF over $N$ is equal to 1, which is equivalent to the Gaussian approximation times $L$ integrated over $ \varphi$ from $-\infty$ to $+\infty$ about its maximum value at $\langle\varphi \rangle$. The  Gaussian approximation is itself clearly only valid when the tails of the distribution are far enough from the extremal values  ($ \varphi = 0$ and 1), which is the case for the three $J=0$ PDFs displayed in Fig.~\ref{fig:OccupanyPDF} (for $\langle\varphi \rangle = 0.31,\,0.5,$ and $0.62$).
For these three cases the Gaussian approximation is extremely accurate.
More generally a PDF calculated using the full approximation for $C_N^L$,  Eq.~\eqref{eq:multapp}, should be accurate over the whole range of $\varphi$ except close to the extremal values. 

In the strong cooperativity limit ($J \to +\infty$), only the empty and full states need be retained in the restricted grand partition function, $\Xi_{+\infty}$, because any state with $0 < N < L$ will necessarily have domain walls and these states will be suppressed via the Boltzmann factor by $\e^{-J}$ (with respect to the fully occupied  state) for each lost nearest-neighbor interaction. This leads to
\be
\Xi_{+\infty} (X;\mu) =  1 + \e^{2 X L},
\ee
where $X = (J+\mu)/2$ is considered to be kept fixed as $J\to + \infty$ and $\mu\to -\infty$. From
$\Xi_{+\infty} (X;\mu)$ we obtain 
\be
\lan N \ran_{+\infty} = \Xi_{+\infty}^{-1} \partial_\mu \Xi_{+\infty}= L \e^{2 X L}/\Xi_{+\infty},
\ee
which allows us to read off the expected results, 
\be
P_{+\infty} (0; \lan\varphi\ran) = \Xi_{+\infty}^{-1} = 1-\lan\varphi\ran
\ee
and
\be
P_{+\infty} (L; \lan\varphi\ran) =  \e^{2 X L} \Xi_{+\infty}^{-1} = \lan\varphi\ran,
\ee
directly from $\Xi_{+\infty}$ [within this approximation the PDF $P_{+\infty} (N; \lan\varphi\ran$) vanishes for all other values of $N$].

For strong anti-cooperativity ($J \ll -1$), we must treat the $L$ odd and even cases differently. In both these cases, the statistical physics is very different from the strong cooperativity ($J \gg 1$) limit where only two states need to be retained. 

\begin{figure}
    \centering
    \includegraphics[scale=0.6]{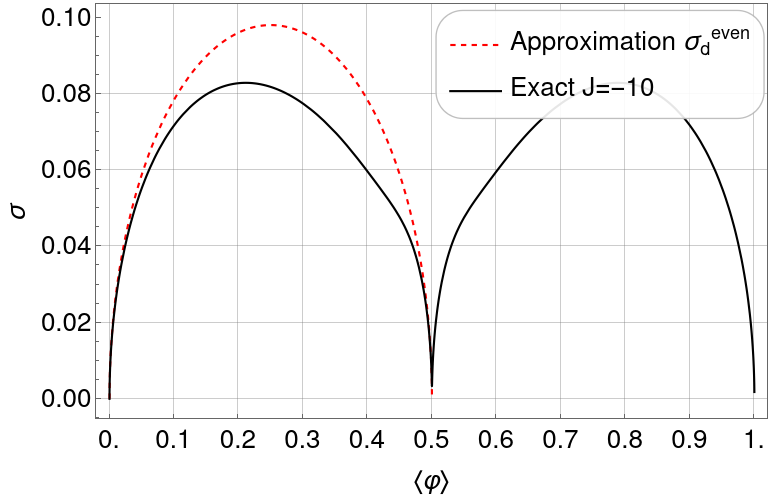}
    \caption{Comparison between the approximation for the very strong anticooperativity regime $\sigma_d^{\rm even}$ and the exact result given by the parametric plot with $\langle \varphi \rangle$ and $\sigma$ (see \textit{even case} in Fig.~\ref{fig:SigmaPhiVsPhi_L14}) for $J = -10$.}
    \label{fig:sigmaeven}
\end{figure}

We start with the simpler $L$ even case for which the system can be approximated by non-overlapping, and therefore non-interacting particle-hole pairs (we choose the convention where the position of the particle-hole pair is determined by the position of the particle making up the particle-hole pair).
The restricted grand partition function for
$0 \le \lan\varphi\ran \le 1/2$
in this particle-hole pair approximation is 
\be
\label{eq:zd}
\Xi_{-\infty}^{\rm even} =   \sum_{N_d=0}^{L/2} D^{L}_{N_d} \e^{\mu N_d},
\ee
where the upper limit on the sum of $L/2$ reflects the maximum number of non-overlapping pairs on a lattice of size $L$ (even) and the multiplicity $D^{L}_{N_d}$ counts the number of distinct ways of putting $N_d$ pairs on a periodic lattice of size $L$. Since calculating $D^{L}_{N_d}$ is a non-trivial combinatorial problem, we simplify matters by considering two disjoint lattices of size $L/2$ consisting of the odd sites for the first and the even sites for the second. If we neglect \emph{mixed} particle-hole pair microstates,  i.e., those that have pairs on both lattices, we obtain the following approximate restricted particle-hole pair grand partition function 
\be
\label{eq:zda}
\Xi_{d}^{\rm even}  =   \sum_{N=0}^{L/2} 2 C^{L/2}_{N_d} \e^{\mu N_d} = 2 \left( 1 + \e^\mu  \right)^{L/2},
\ee
where $C^{L/2}_{N_d}$ is the usual binomial coefficient that counts the number of distinct ways of placing $N_d$ particles on a lattice of size $L/2$. In passing from Eq.~\eqref{eq:zd} to Eq.~\eqref{eq:zda}, we have 
replaced $D^{L}_{N_d}$ by $2 C^{L/2}_{N_d}$ and therefore
simplified the problem to that of two independent non-interacting \emph{quasi-particle}  (i.e., particle-hole pair) systems [cf. Eq.~\eqref{eq:zo})]. 
Following the above discussion for the $J=0$ case, we can immediately conclude that $\langle\varphi\rangle =\langle N_d/L \rangle \approx 
\langle\varphi\rangle_d^{\rm even}$, where
\be
\langle\varphi\rangle_d^{\rm even} = 
(L \Xi_d)^{-1}\partial_\mu \Xi_d^{\rm even} = 1/\left[2\left( 1 + \e^{-\mu}\right)  \right]
\ee
and therefore $\sigma_{-\infty}^{\rm even} \approx \sigma_d^{\rm even}$,
where
 \be
 \sigma_d^{\rm even} =  \sqrt{L^{-1} \partial_\mu \langle \varphi \rangle_d^{\rm even}} = 
 L^{-1/2}  \sqrt{\langle \varphi \rangle - 2 \langle \varphi \rangle^2}.
 \ee
The replacement of $D^{L}_{N_d}$
by $2 C^{L/2}_{N_d}$ is exact for the one particle-hole pair state ($N_d =1$) and at half-filling, 
($N_d = L/2$), two cases for which there are no allowed mixed-states (in the latter case because of the high filling of pairs).
Although we over-count the (unique) empty state by a factor of 2 (in order to put $\Xi_{d}^{\rm even}$ in the free particle form)  
and under-count the number of particle-hole pair states for $1 < N_d < L/2$, for which there are allowed mixed-states, the physics captured by the approximation $\sigma_{d}^{\rm even}$ is sufficiently faithful to that of the original system to be able to account well for the main quantity of interest, namely the standard deviation in the strong anti-cooperativity limit for $L$ even 
(see Fig.~\ref{fig:sigmaeven}).

Inspection of Fig.~\ref{fig:SigmaPhiVsPhi_L14} and Fig.~\ref{fig:sigmaeven} shows that the approximate form $\sigma_d^{\rm even}$ accounts well for the exact behavior, going to 0 at zero and HF and reaching a maximum at 1/4 filling, not far from the exact maximum position. Because of the approximations inherent in obtaining
$\sigma_d^{\rm even}$, however, the approximate result cannot capture the weak lack of symmetry about 1/4 filling, i.e., the observed skewness towards lower fillings seen in  Figs.~\ref{fig:SigmaPhiVsPhi_L14} and \ref{fig:sigmaeven}.
This skewness can be seen
even more clearly  in the PDFs for $L=14$ and 
$\langle \varphi \rangle = 1/3$
presented in Fig.~\ref{fig:PDF-strong-anticooperqtivity}. 
It arises because for $1 < N_d < L/2$  the approximation $2 C^{L/2}_{N_d}$  underestimates the exact particle-hole pair multiplicity $D^{L}_{N_d}$ more severely for lower fillings than for higher ones (below HF). The reason being that at higher fillings it becomes more and more difficult to have allowed mixed states. 

The roughly Gaussian shape for the PDF with $L=14$ presented in Fig.~\ref{fig:PDF-strong-anticooperqtivity} for $\lan \varphi \ran = 1/3$ can be understood using the approximate  particle-hole pair approach that maps the problem onto free quasi-particles (particle-hole pairs). In this case we can adapt the results for the $J=0$ case to estimate the particle-hole pair PDF for $J \to -\infty$ and $L$ even,
 \be
 P_d^{\rm even}(N; \lan \varphi \ran) \equiv \frac{C^{L/2}_{N_d} \e^{\mu_d^{\rm even} N_d}}{\Xi_d^{\rm even}}, 
 \ee
 using the Gaussian approximation:
  \be
 P_d^{\rm even}(N; \lan \varphi \ran) 
 \approx  
 \frac{\exp 
 \left[ -\frac{(\varphi - \lan \varphi \ran)^2}{2 (\sigma_d^{\rm even})^2}  
 \right]}{{(2 \pi)^{1/2} \sigma_d^{\rm even} L}}, 
 \ee
where we have used
\be
\Xi_d^{\rm even}(\lan \varphi \ran) = 
\frac{2}{( 1-2\lan \varphi \ran)^{L/2}},
\ee
and 
\be
\mu_d^{\rm even}(\lan \varphi \ran) = \ln \left( \frac{2 \lan \varphi \ran}{1-2 \lan \varphi \ran} \right).
\ee
The  Gaussian PDF approximation is very accurate compared with $P_d^{\rm even}(N; \lan \varphi \ran)$ for $L=14$  and $\lan \varphi \ran = 1/3$, although $P_d^{\rm even}(N; \lan \varphi \ran)$ itself underestimates the exact maximum by about 30\% and cannot capture the skewness observed in the exact results (see Fig.~\ref{fig:PDF-strong-anticooperqtivity}).
The particle-hole pair approximation does reproduce accurately, however, the width of the exact PDF.

In the strong anti-cooperativity limit ($J \to - \infty$) for $L$ odd, the system cannot be approximated by non-overlapping particle-hole pairs at HF. We need to account for a defect in order to approach HF. 
The restricted grand partition function for
$0 \le \lan\varphi\ran \le 1/2$
in this particle-hole pair approximation is 
\begin{equation}
\label{eq:zdodd}
\Xi_{-\infty}^{\rm odd} =   \sum_{N_d=0}^{(L-3)/2} O^{L}_{N_d} \e^{\mu N_d} + 
L \e^{\mu (L-1)/2} + 
L \e^{ - |J| + \mu (L+1)/2},
\end{equation}
where $(L-1)/2$ is the maximum number of non-overlapping pairs on a lattice of odd size $L$ and the multiplicity $O^{L}_{N_d}$ counts the number of distinct ways of putting $N_d$ pairs on a periodic lattice of odd size $L$. 
The last two terms are the defect contributions, an extra hole or particle not making up a particle-hole pair that costs an energy $-J$ for an extra particle and that leads to a multiplicity $L$ counting the number of ways to place an extra hole or particle on a lattice made up of $(L-1)/2$ pairs in sequence.

Since calculating $O^{L}_{N_d}$ is a non-trivial combinatorial problem, we follow a procedure similar to the one used in the even case and simplify the problem by considering two disjoint lattices of size $(L+1)/2$. If we neglect \emph{mixed} particle-hole pair microstates for $0 \le N_d \le (L-3)/2$,  i.e., those that have pairs on both lattices, we obtain the following approximate restricted particle-hole pair grand partition function 
\begin{equation}
\Xi_{d}^{\rm odd} =   \sum_{N_d=0}^{(L-3)/2} 2 C^{(L+1)/2}_{N_d} \e^{\mu N_d} +
L \e^{\mu (L-1)/2} + 
L \e^{ - |J| + \mu (L+1)/2},
\end{equation}
which can be evaluated by completing the binomial sum and then subtracting the added terms:
\begin{equation}\label{eq:zdaodd}
\Xi_{d}^{\rm odd} =   2 \left( 1 + \e^\mu  \right)^{(L+1)/2} 
 - \e^{\mu (L-1)/2}  
 - \e^{ \mu (L+1)/2} ( 2 - L \e^{ - |J|} )
\end{equation}
where $C^{(L+1)/2}_{N_d}$ counts the number of distinct ways of placing $N_d$ particles on a lattice of size $(L+1)/2$.  $\Xi_{d}^{\rm odd}$ correctly counts the multiplicity of the $(L\pm 1)/2$ states by construction. 
Although this approximation overcounts the one particle-hole pair state by 1 ($L+1$ instead of $L$) and overcounts the empty state by a factor of 2, we believe that it captures the essential physics of the strong anti-cooperativity case for $L$ odd.
The above approximation, $\Xi_{d}^{\rm odd}$, can be used to calculate
$\langle\varphi\rangle_{d}^{\rm odd} = 
(L \Xi_{d}^{\rm odd})^{-1}\partial_\mu \Xi_{d}^{\rm odd}$
and  $\sigma_{d}^{\rm odd} =  \sqrt{L^{-1} \partial_\mu \langle \varphi \rangle_{d}^{\rm odd}}$, an approximate results that could then be compared with the exact ones. We will not, however, pursue this approach any further here.

\subsection{Infinite range (IR) model}\label{sec:IR-model}
The mapping between the SR and IR Lattice Gas can be established by performing a perturbative cumulant expansion of the grand partition function:
\bea
\Xi & = & \sum_{\left\{\varphi_i\right\}} \e^{-H_0-H_{\rm int}}  \nonumber \\
  & = &  \Xi_0 \langle \e^{-H_{\rm int}} \rangle_0  \nonumber \\
  & \approx  &  \Xi_0 \e^{-\langle H_{\rm int}\rangle_0},
\eea
where the last line gives the first order cumulant expansion. The subscript 0 indicates a statistical average with respect to the non-interacting (Hill-Langmuir) model (ideal Lattice Gas).
By calculating $\langle H_{\rm int}\rangle_0$,
we obtain $\langle H_{\rm int}\rangle_0 =
-J_{\rm SR} L \langle \varphi \rangle_0^2 $ for the SR model and
$-\frac{1}{2}J_{\rm IR} L (L-1) \langle \varphi \rangle_0^2$ for the IR model, results that suggest the following mapping between the two models:
\be
J_{\rm SR} \leftrightarrow \frac{1}{2} J_{\rm IR}(L-1).
\label{EQ:mapping}
\ee
This mapping is exact to lowest order in the coupling constant.

We adopt the following strategy to avoid carrying out an explicit calculation of the standard deviation $\sigma$ directly within the cumulant approximation, which is cumbersome.
The weak coupling results for the SR model can be found by expanding the exact parametric results for $\langle \varphi \rangle (\mu)$ and the variance $v(\mu) = \sigma^2(\mu)$ in powers of $J$.
To first order in $J_{\rm SR} $ we find
\be
\left(\frac{\sigma_0}{\sigma_{\rm SR} }\right)^2 = 1- 2 J_{\rm SR} L \sigma_0^2
+ {\cal O} (J_{\rm SR} ^2) = 1- 2 J_{\rm SR} \sigma_\infty^2
+ {\cal O} (J_{\rm SR} ^2).
\ee
We can then use the above mapping between the SR and LR models to get a result that is exact to first order in $J_{\rm IR} $ for the IR model:
\be
\left(\frac{\sigma_0}{\sigma_{\rm IR} }\right)^2 = 1- L(L-1) J_{\rm IR} \sigma_0^2
+ {\cal O} (J_{\rm IR}^2) = 1- (L-1) J_{\rm IR} \sigma_\infty^2
+ {\cal O} (J_{\rm IR}^2) .
\ee

One can check the above method  for SR model at half-filling:
\be
\left[1- \left(\frac{\sigma_0}{\sigma_{\rm SR} }\right)^2 \right]  =
\frac{1}{2} J_{\rm SR}
+ {\cal O} (J_{\rm SR}^2 )   \qquad \text{(half-filling).}
\ee

It would be useful to find a good interpolation formula for $\sigma$ as an explicit function of $L$, $J$, and $\langle \varphi \rangle$ for both the SR and IR  models.  
A reasonably good interpolation scheme can be set up by recognizing that the exact result for the standard deviation at half-filling for small $J$  is to a very good approximation,
\be
\left(\frac{\sigma_0}{\sigma_{\rm SR} }\right)^2
= \exp\left(- 2 J_{\rm SR} \sigma_\infty^2\right) \times
 \left[ 1 + {\cal O} \left(\left[\sigma_\infty^2 J_{\rm SR}\right]^L\right) \right]
\qquad \text{(half-filling)}
\ee
(where $\sigma_\infty^2 = 1/4$ at HF), which shows that it's best to use an exponential resummation of the small $J$ perturbation expansion.

Comparing the exact SR results with the above approximation shows that an even better approximation is to take the 5/2 root mean (a heuristic choice),
\be
\left(\frac{\sigma_0}{\sigma_{\rm SR} }\right)^2
\approx
\left[ \exp(- 5 J_{\rm SR} \sigma_\infty^2)+ L^{-5/2} \right]^{2/5},
\ee
to get a smooth interpolation.
The above 5/2 root mean square average leads to our best approximation for
$\sigma_{\rm SR}$:
\be
\sigma_{\rm SR}
\approx
\frac{\sigma_0}
{\left[ \exp(- 5 J_{\rm SR} \sigma_\infty^2)+ L^{-5/2} \right]^{1/5}}.
\label{EQ:sigmaaprox}
\ee
This is a convenient form because the dependence on
$\langle \varphi\rangle$ (through $\sigma_\infty$), $J_{\rm SR}$, and $L$ is made explicit
(since $\sigma_0=L^{-1/2} \sigma_\infty$). 
The crossover to the large $J$ saturation limit is centered on
\be
J_{\rm co} \equiv \ln(L)/(2\sigma_\infty^2).
\ee
$J_{\rm co}$ depends on $L$ and filling $\langle \varphi\rangle$, increasing slowly with $L$ and reaching a minimum at half-filling for fixed $L$.
The convergence of $\sigma$ as a function of $\langle\varphi\rangle$ to the large $J$ saturation limit is uniform only for
$J < J_{\rm co}$.

\begin{figure}[ht]
\centering
    \includegraphics[width=86mm]{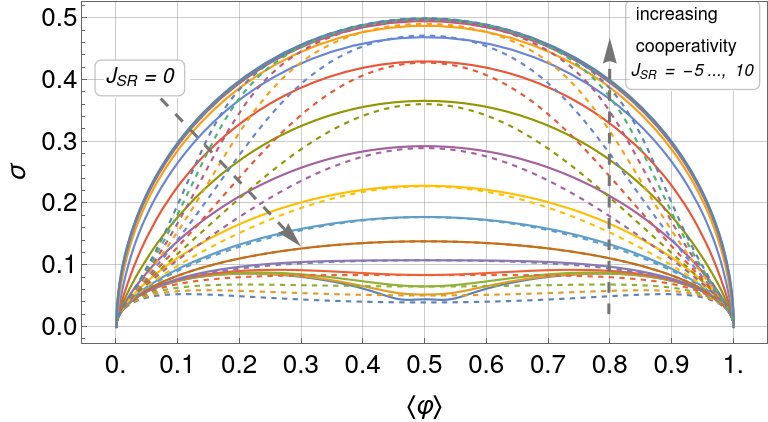}
    \caption{Comparison between the exact result for the standard deviation (solid lines) and the approximate expression (dashed lines) that depends explicitly on $\langle \varphi \rangle$, $J$, and $L$ (Eq.~(\ref{EQ:sigmaaprox})).}  
    \label{fig:sigmaaprox}
\end{figure}

The approximation \eqref{EQ:sigmaaprox} 
is very accurate for all fillings (except possibly very close to zero and full-filling) as long as $L^{-5/2} \ll 1$ (the case for $L \approx 10$) and $J < J_{\rm co}$ (see Fig.~\ref{fig:sigmaaprox}). 
For $L^{-5/2} \ll 1$ and
$J > J_{\rm co}$, this approximation is  accurate only within a window centered on half-filling.  Because of the factor $\sigma_\infty^2$ in the argument of the exponential, convergence to $\sigma_\infty$ is fastest at HF and becomes slower and slower as moves away from HF to zero and full filling. For $L \approx 10$, the window of accuracy for  $J<2$ is
$0.2 < \langle\varphi\rangle < 0.8$, which encompasses the experimental data window.

The interpretation of the above results is in accordance with our previous understanding: as long as the system is far enough from  zero and full-filling, increasing $J$ leads to a flattening of the Gaussian PDF with $\sigma$ growing exponentially with $J$ until the system reaches the crossover region, $J \approx J_{\rm co}$, before saturating at $\sigma_\infty$ for $J \gg J_{\rm co}$.
When the system reaches the crossover region the Gaussian form is no longer accurate because one (or two) of the wings starts to touch one (or two) of the boundaries, before morphing into a bimodal (two-state) system.

The corresponding result for the IR case can be obtained using the mapping \eqref{EQ:mapping}. 
We note in passing that an approximate Gaussian effective free energy approach can be developed for the IR model along the lines of the one developed above for the non-interacting (Hill-Langmuir) model. This result can be used to find an extended Gaussian approximation for the IR model,
\be
 P_IR(N; \lan \varphi \ran) \approx  
 \frac{\exp \left[ -\frac{(\varphi - \lan \varphi \ran)^2}
 {2 \sigma_{\rm IR}^2} \right]}
 {{(2 \pi)^{1/2} \sigma_{\rm IR} L}},
 \ee
where the variance $\sigma_{\rm IR}$ can be approximated using the approximation for the SR model 
\eqref{EQ:sigmaaprox}
and the mapping \eqref{EQ:mapping}.

\subsection{System size, fluctuations, and experimental precision \label{sec:syssize}}

By proposing a pragmatic operational definition, we clarify here what we mean by a \textit{small system}. 
Before doing so we note that if we were to attempt to qualify the size of an adsorption system by the ratio between the number of binding sites, $L$, and the size of a correlated domain (twice the correlation length), we would only be characterizing the Hill-Langmuir nature of the fluctuations, occurring when $2\xi/L \ll 1$, or
the bimodal nature of the fluctuations, occurring when $2\xi/L \gg 1$, and not the feasibility of extracting the cooperativity from the experimentally measured relative occupancy fluctuations at equilibrium.

We propose here that a system be considered a \textit{small} cooperative one if fluctuations, as quantified by the standard deviation of relative occupancy, are within experimental resolution and can be used to extract, using the theoretical framework that we have developed, the amplitude of the interaction strength, $J$, and that this strength be within the usual biophysical range, $0 \leq J \leq J_{\rm max} \approx 10$, for adsorption systems.

\begin{figure}
\centering
\begin{subfigure}{0.45\textwidth}
\includegraphics[width=0.9\textwidth]{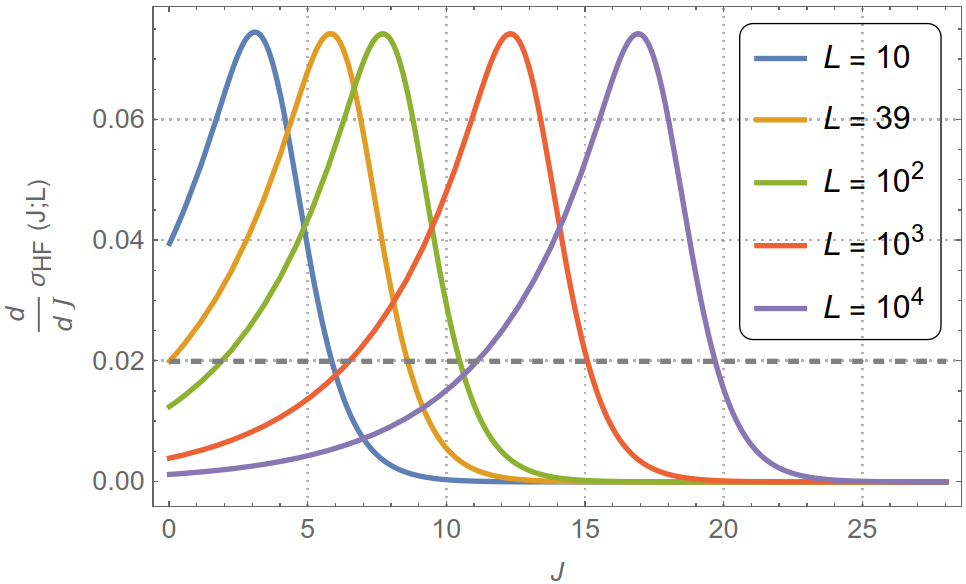}
\caption{\label{fig:derivative_of_SD}}
\end{subfigure}
\begin{subfigure}{0.45\textwidth}
\includegraphics[width=0.9\textwidth]{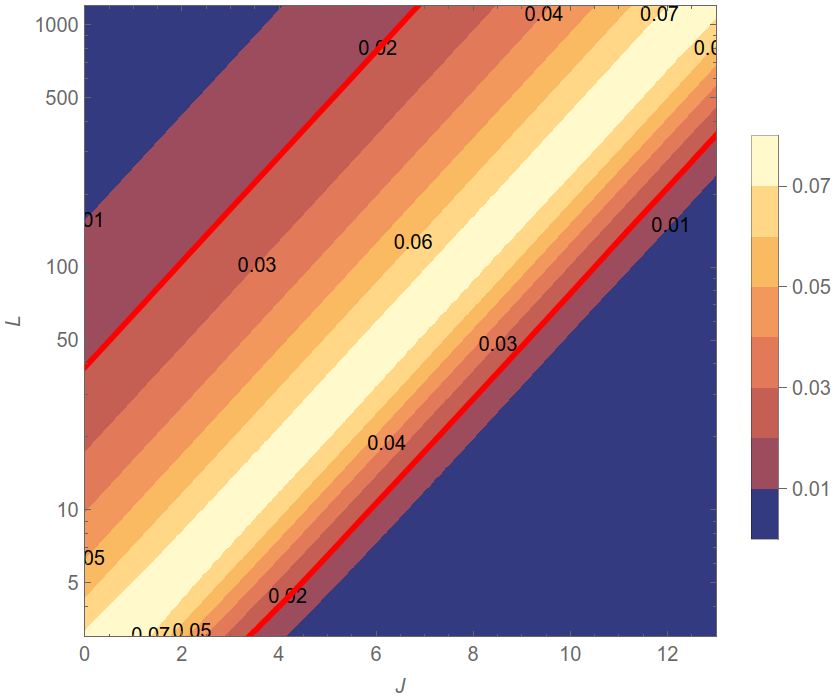}
\caption{}
\label{fig:LvsJ_heatmap}
\end{subfigure}
\caption{
\label{fig:L-J-fluctuations} (a) Derivative of the standard deviation at half-filling, 
$\sigma_\text{HF}'(J;L) \equiv \text d\sigma_\text{HF}/\text dJ$,
plotted for systems of size $L=10,\, 39,\, 10^2,\,10^3,\,10^4$, respectively. 
The horizontal dashed gray line represents the threshold value $\Delta\sigma=0.02$.
(b) Contour plot of  $\sigma_\text{HF}'(J;L)$ for positive values of $J$ and $L\geq 3$.
The red contour lines correspond to $\sigma_\text{HF}'(J;L) = \Delta\sigma=0.02$.}
\end{figure}

To quantify our capacity to resolve $J$ theoretically, 
we therefore propose the following criterion:  the derivative of $\sigma$ with respect to $J$,
$\sigma_\text{HF}'(J;L)$, \emph{must be larger than a given value} $\Delta\sigma$, whose choice depends on the measurement precision, over a $J$ range starting at 0.
In what follows, for the application to the BFM, we shall choose $\Delta\sigma = 0.02$, which is compatible with the experimental error for the data provided in Section \ref{sec:data}.
This choice is motivated by comparing the parametric plots in Fig.~\ref{fig:SigmaPhiVsPhi+L13+L14} to the experimental data (see Fig.~\ref{fig:std_vs_phi_EXP}). 
One can estimate the coupling $J$ from the data when these fall in regions of the parametric plots where the lines corresponding to different values of $J$ are spaced far enough apart compared to the experimental error.
The highest resolution for positive values of $J$, and fixed $L\geq 3$, is achieved at half-filling (HF), and for a value 
$J^* \approx 2 \ln (L/2)$ 
that maximizes the derivative of the standard deviation. 
These maxima correspond to the peaks of the curves in Fig.~\ref{fig:derivative_of_SD} where we plot $\sigma_\text{HF}'(J;L)$ as a function of $J$ for different but fixed values of $L$ and  compare it with the threshold value 
$\Delta\sigma=0.02$.

We observe that for the choice of $\Delta\sigma=0.02$, $L^* \approx 39$ is the maximum system size for which there is a feasibility window ($0 < J < 8$) starting at 0 for extracting the cooperativity from the experimental data. This threshold value $L^*$
can be recovered analytically by using the approximation
Eq.~\eqref{EQ:sigmaaprox} for the standard deviation, which is highly accurate near half-filling provided $L \gg 1$, to find 
$\sigma_\text{HF}'(J=0; L) \approx 1/(8 L^{1/2})$. 
One can then solve $\sigma_\text{HF}'(J=0; L) = \Delta\sigma = 0.02$ to obtain $L^* = 1/(64 \Delta\sigma^2)$.
The same approximation leads to
$\sigma_\text{HF}''(J=0; L) \approx 1/(32 L^{1/2})$. 
These approximations show that both the first and second derivatives of $\sigma_\text{HF}$ at zero coupling ($J=0$) decrease as 
$L^{-1/2}$. 
For high coupling, $J \gg J_\text{max}$, the system assumes a bimodal character and the standard deviation saturates at the infinite coupling limit.

We conclude, based on our definition and our choice of $\Delta\sigma$, that a  system is  \textit{small} if $L \leq 39$. We also observe that for any system of finite size with $L > 39$, there is always a feasibility window for sufficiently strongly cooperative systems, 
$J_{\rm min} < J < J_{\rm max}$, characterized by a peak at a value of $J = J^*$ that increases logarithmically with increasing $L$, but with $J_{\rm min} > 0$.
Consequently, for $L \gg 39$, the standard deviation at HF is insensitive to $J$ both at low coupling, where the system acts as a Hill-Langmuir one, and high coupling, where the system becomes bimodal.

The (semi-log) contour plot in Fig.~\ref{fig:LvsJ_heatmap} gives the isolines of $\sigma_\text{HF}'(J)$ as a function of the interaction potential $J$ and the system size $L$, where the latter, for the sake of simplicity, is treated as a continuous variable.
Recall the asymptotic behavior of the correlation length at half-filling for large values of the interaction potential 
\cite{Godreche2000response}: $J \sim 2 \ln\left( 2 \xi_\text{HF}\right)$.
Hence, if we were to plot the logarithm of $\xi_\text{HF}$ instead of $J$ on the horizontal axis, we would get a qualitatively equivalent contour diagram.
The two red contour lines correspond to the constant value $\Delta\sigma=0.02$. 
The region between them indicates the set of parameters $J$ and $L$ for which $\sigma_\text{HF}'(J)>\Delta\sigma$.
We finally conclude that a system of size $L=13$, like the one considered in the main text to describe the BFM, is well within the feasibility window for small systems.

\subsection{Effective Hill coefficient}\label{ssec:methods-Hillcoeff}
Following the analysis in \cite{owen2023size} on the logarithmic sensitivity of kinetics schemes, we can also
quantify cooperative behavior by introducing an effective Hill coefficient which, for our system, takes the following form:
\begin{equation}\label{eq:HillCoefficient}
H_\textrm{eff}= \frac{\text d}{\text d \mu}\ln\left(\frac{\langle\varphi \rangle}{1-\langle\varphi\rangle}\right) 
= \frac{\sigma^2 L}{\langle \varphi \rangle \left(1-\langle \varphi\rangle\right)}.
\end{equation}
Here, we have exploited the relation $\partial_\mu \langle \varphi\rangle = \sigma^2 L$.

\begin{figure}
    \centering
    \includegraphics[width=0.5\textwidth]{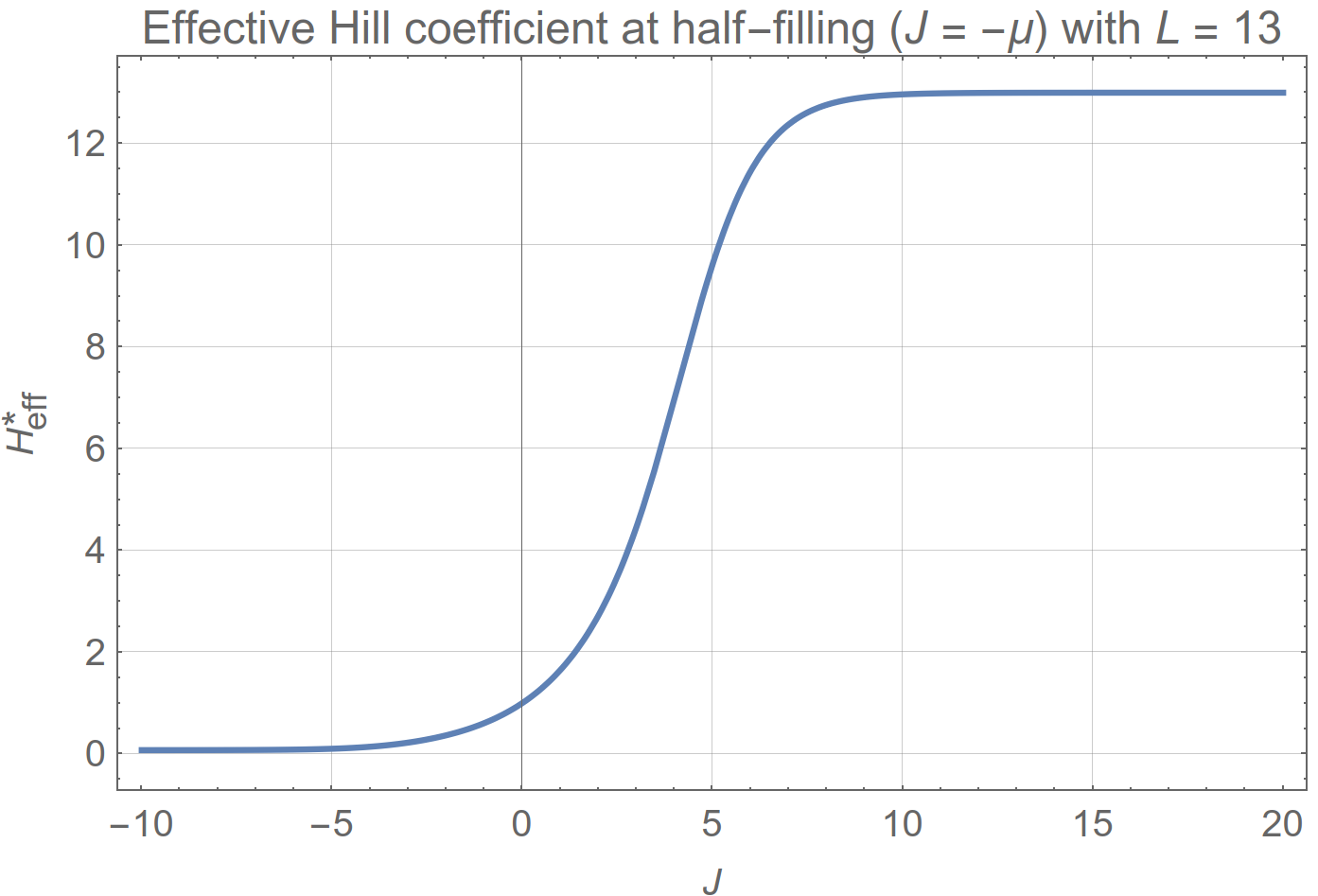}
    \caption{The effective Hill coefficient at half-filling from Eq. (\ref{eq:HillCoefficientMax}) as a function of the interaction potential $J$ for a system of size $L=13$.}
    \label{fig:EffectiveHillCoefficient}
\end{figure}

The effective Hill function defined above takes on a maximal value $H_\text{eff}^*$ at half-filling:
\begin{equation}\label{eq:HillCoefficientMax}
    H_\textrm{eff}^*(J) = 4\sigma_\text{HF}^2L,
\end{equation}
where $\sigma_\text{HF}^2$ is the square of the relative standard deviation at half-filling whose explicit dependency on $J$ is captured by Eq.~(\ref{eq:sigmaHF}). 
This expression tends rapidly to zero for negative values of $J$ (anticooperativity).
Moreover, for $J=0$, the coefficient equals 1, as expected for the Hill-Langmuir case.
Finally, by increasing the interaction potential, $\sigma_\text{HF}$ tends to $1/2$, and so $H^*_\text{eff}\to L$.
This result is in agreement with the rule that at equilibrium, the Hill coefficient is expected not to exceed the maximal number of ligands that can bind simultaneously, which, in the case of an adsorption process, corresponds to the number of binding sites, $L$. 
Fig.~\ref{fig:EffectiveHillCoefficient} shows the sigmoidal allure of the Hill coefficient, bounded from above by the system size and below by 0.

It should be noted, however, that in \cite{owen2023size} it has been demonstrated that this upper bound for the effective Hill coefficient may be surpassed when the system is out-of-equilibrium. 
In such cases, the new limit is determined not by the size of the lattice, but by the number of states of the kinetic scheme whose exit rates increase with the concentration of the ligands.

\subsection{Supplemental plots}
\begin{figure}[ht]
\centering
\includegraphics[width=86mm]{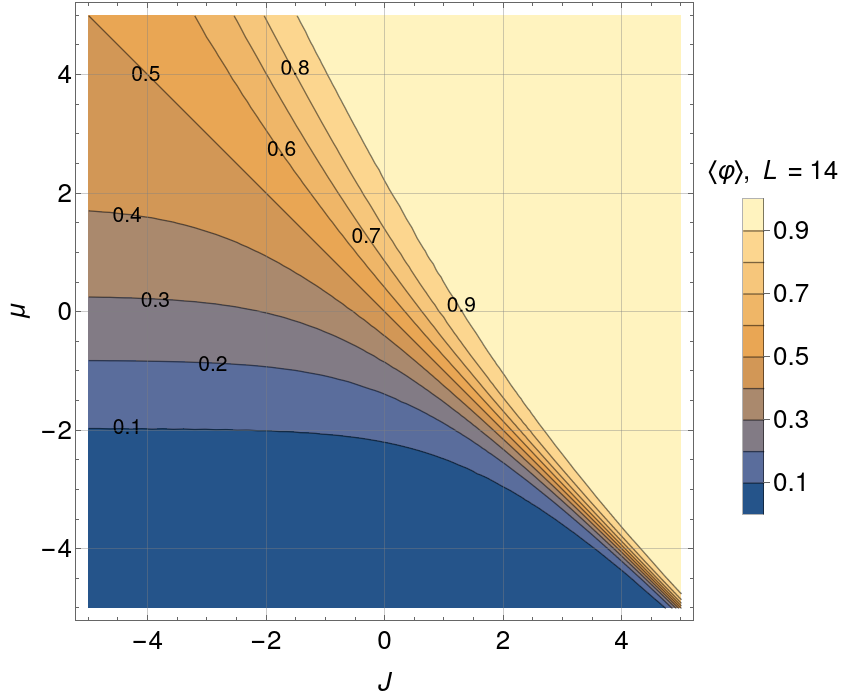}
\caption{\label{fig:Phi-L14} Contour plot of the mean relative occupancy in equilibrium, $\langle\varphi\rangle$, as a function of the dimensionless interaction potential $J$ and chemical potential $\mu$ for a system of size $L = 14$.}
\end{figure}
\begin{figure}[ht]
\centering
	\includegraphics[width=86mm]{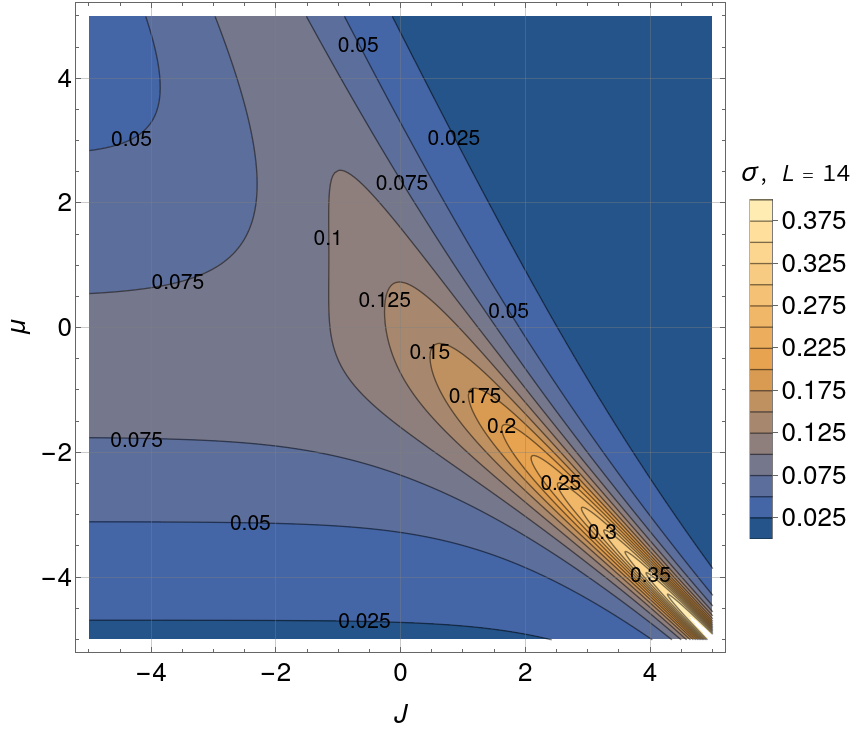}
	\caption{\label{fig:SigmaPhi-L14} Contour plot of the standard deviation of the relative occupancy at equilibrium, $\sigma = \sqrt{\langle \varphi^2 \rangle - \langle \varphi \rangle^2}$, as a function of the dimensionless interaction potential $J$ and chemical potential $\mu$ for a system of size $L=14$.}
\end{figure}

\clearpage
\subsection{Experimental data}

\begin{table}[h!]
\caption{Data of average occupancy and standard deviation corresponding to Fig. \ref{fig:std_vs_phi_EXP}. 
Data are obtained by averaging over several traces \cite{Nord2017Catch, Perez-Carrasco2022}.
As there are not enough traces for good statistics (from 30 to 40), data is also averaged over several time points at equilibrium.}
\centering
\begin{tabular}{c|cc}
\begin{tabular}[c]{@{}c@{}} Bead size\\ (drag)\end{tabular} & \multicolumn{1}{c|}{$\langle \varphi \rangle$} & $\sigma$          \\ \hline
300 nm                                                     & $0.29 \pm 0.03$                              & $0.17 \pm 0.02$ \\
                                                           & $0.355 \pm 0.006$                              & $0.153 \pm 0.011$ \\
                                                           & $0.41 \pm 0.02$                              & $0.15 \pm 0.02$ \\ \hline
300 nm + Glycerol                                          & $0.461 \pm 0.005$                              & $0.175 \pm 0.007$ \\ \hline
500 nm                                                     & $0.56 \pm 0.02$                              & $0.18 \pm 0.02$ \\
\multicolumn{1}{l|}{}                                      & $0.592 \pm 0.011$                              & $0.21 \pm 0.02$ \\
\multicolumn{1}{l|}{}                                      & $0.67 \pm 0.02$                              & $0.18 \pm 0.04$ \\ \hline
500 nm + Glycerol                                          & $0.705 \pm 0.005$                              & $0.158 \pm 0.005$ \\ \hline
1300 nm                                                    & $0.74 \pm 0.03$                              & $0.13 \pm 0.02$ \\
\multicolumn{1}{l|}{}                                      & $0.80 \pm 0.03$                              & $0.153 \pm 0.012$ \\
\multicolumn{1}{l|}{}                                      & $0.828 \pm 0.006$                              & $0.18 \pm 0.04$ \\
\multicolumn{1}{l|}{}                                      & $0.826 \pm 0.011$                              & $0.161 \pm 0.009$
\end{tabular}

\end{table}

\begin{table}[h!]
\caption{Data of probability distribution corresponding to Fig.\ref{fig:OccupancyPDF_EXP}. Data are obtained by averaging over several time points at equilibrium the state of all traces.}
\centering
\begin{tabular}{c|ccc}
                                                        & \multicolumn{3}{c}{\cellcolor[HTML]{EFEFEF}Bead size}                                               \\ \cline{2-4} 
\cellcolor[HTML]{FFFFFF}                                & \multicolumn{1}{c|}{300 nm}            & \multicolumn{1}{c|}{500 nm}            & 1300 nm           \\ \hline
\rowcolor[HTML]{EFEFEF} 
\multicolumn{1}{l|}{\cellcolor[HTML]{EFEFEF}\# Stator units} & \multicolumn{3}{c}{\cellcolor[HTML]{EFEFEF}{\color[HTML]{333333} Probability}}                      \\ \hline
0                                                       & \multicolumn{1}{c|}{$0.12 \pm 0.04$}   & \multicolumn{1}{c|}{$0.005 \pm 0.014$} & $0.01 \pm 0.20$   \\
1                                                       & \multicolumn{1}{c|}{$0.09 \pm 0.04$}   & \multicolumn{1}{c|}{$0.005 \pm 0.013$} & $0$               \\
2                                                       & \multicolumn{1}{c|}{$0.11 \pm 0.05$}   & \multicolumn{1}{c|}{$0$}               & $0$               \\
3                                                       & \multicolumn{1}{c|}{$0.15 \pm 0.04$}   & \multicolumn{1}{c|}{$0.01 \pm 0.02$}   & $0$               \\
4                                                       & \multicolumn{1}{c|}{$0.15 \pm 0.04$}   & \multicolumn{1}{c|}{$0.05 \pm 0.02$}   & $0$               \\
5                                                       & \multicolumn{1}{c|}{$0.22 \pm 0.04$}   & \multicolumn{1}{c|}{$0.04 \pm 0.03$}   & $0$               \\
6                                                       & \multicolumn{1}{c|}{$0.13 \pm 0.03$}   & \multicolumn{1}{c|}{$0.17 \pm 0.05$}   & $0$               \\
7                                                       & \multicolumn{1}{c|}{$0.033 \pm 0.014$} & \multicolumn{1}{c|}{$0.29 \pm 0.10$}   & $0.002 \pm 0.011$ \\
8                                                       & \multicolumn{1}{c|}{$0$}               & \multicolumn{1}{c|}{$0.13 \pm 0.04$}   & $0.12 \pm 0.04$   \\
9                                                       & \multicolumn{1}{c|}{$0$}               & \multicolumn{1}{c|}{$0.13 \pm 0.04$}   & $0.14 \pm 0.03$   \\
10                                                      & \multicolumn{1}{c|}{$0$}               & \multicolumn{1}{c|}{$0.05 \pm 0.02$}   & $0.28 \pm 0.07$   \\
11                                                      & \multicolumn{1}{c|}{$0$}               & \multicolumn{1}{c|}{$0.04 \pm 0.02$}   & $0.25 \pm 0.06$   \\
12                                                      & \multicolumn{1}{c|}{$0$}               & \multicolumn{1}{c|}{$0.06 \pm 0.02$}   & $0.18 \pm 0.05$   \\
13                                                      & \multicolumn{1}{c|}{$0$}               & \multicolumn{1}{c|}{$0.03 \pm 0.02$}   & $0.02 \pm 0.03$  
\end{tabular}
\end{table}

\begin{table}[h!]
\centering
\caption{}
\begin{tabular}{c|ccc}
                          & \multicolumn{3}{c}{\cellcolor[HTML]{EFEFEF}$\mu$ estimated from $\langle \varphi  \rangle$}                                      \\ \hline
\rowcolor[HTML]{EFEFEF} 
Bead size                 & \multicolumn{1}{c|}{\cellcolor[HTML]{EFEFEF}$J = 1$} & \multicolumn{1}{c|}{\cellcolor[HTML]{EFEFEF}$J = 2$} & $J = 3$            \\ \hline
                          & \multicolumn{1}{c|}{$-1.56 \pm 0.08$}                & \multicolumn{1}{c|}{$-2.35 \pm 0.05$}                & $-3.21 \pm 0.03$   \\
                          & \multicolumn{1}{c|}{$-1.37 \pm 0.02$}                & \multicolumn{1}{c|}{$-2.222 \pm 0.011$}              & $-3.14 \pm 0.01$   \\
\multirow{-3}{*}{300 nm}  & \multicolumn{1}{c|}{$-1.23 \pm 0.04$}                & \multicolumn{1}{c|}{$-2.14 \pm 0.03$}                & $-3.08 \pm 0.02$   \\ \hline
300 nm + Glycerol         & \multicolumn{1}{c|}{$-1.096 \pm 0.012$}              & \multicolumn{1}{c|}{$-2.058 \pm 0.007$}              & $-3.035 \pm 0.004$ \\ \hline
                          & \multicolumn{1}{c|}{$-0.85 \pm 0.05$}                & \multicolumn{1}{c|}{$-1.91 \pm 0.03$}                & $-2.95 \pm 0.02$   \\
                          & \multicolumn{1}{c|}{$-0.77 \pm 0.03$}                & \multicolumn{1}{c|}{$-1.86 \pm 0.02$}                & $-2.92 \pm 0.01$   \\
\multirow{-3}{*}{500 nm}  & \multicolumn{1}{c|}{$-0.56 \pm 0.06$}                & \multicolumn{1}{c|}{$-1.73 \pm 0.04$}                & $-2.84 \pm 0.02$   \\ \hline
500 nm + Glycerol         & \multicolumn{1}{c|}{$-0.46 \pm 0.02$}                & \multicolumn{1}{c|}{$-1.67 \pm 0.01$}                & $-2.80 \pm 0.01$   \\ \hline
                          & \multicolumn{1}{c|}{$-0.32 \pm 0.11$}                & \multicolumn{1}{c|}{$-1.58 \pm 0.07$}                & $-2.75 \pm 0.04$   \\
                          & \multicolumn{1}{c|}{$-0.12 \pm 0.011$}               & \multicolumn{1}{c|}{$-1.45 \pm 0.07$}                & $-2.67 \pm 0.05$   \\
\multirow{-3}{*}{1300 nm} & \multicolumn{1}{c|}{$0.012 \pm 0.032$}               & \multicolumn{1}{c|}{$-1.37 \pm 0.02$}                & $-2.614 \pm 0.013$
\end{tabular}
\end{table}

\begin{table}[h!]
\centering
\caption{}
\begin{tabular}{c|ccc}
                                  & \multicolumn{3}{c}{\cellcolor[HTML]{EFEFEF}$\mu$ estimated from $\sigma$}                                                      \\ \hline
\rowcolor[HTML]{EFEFEF} 
\cellcolor[HTML]{EFEFEF}Bead size & \multicolumn{1}{c|}{\cellcolor[HTML]{EFEFEF}$J = 1$} & \multicolumn{1}{c|}{\cellcolor[HTML]{EFEFEF}$J = 2$} & $J = 3$          \\ \hline
                                  & \multicolumn{1}{c|}{$-1.4 \pm 0.4$}                  & \multicolumn{1}{c|}{$-2.55 \pm 0.11$}                & $-3.48 \pm 0.06$ \\
                                  & \multicolumn{1}{c|}{$-1.61 \pm 0.21$}                & \multicolumn{1}{c|}{$-2.63 \pm 0.09$}                & $-3.53 \pm 0.05$ \\
\multirow{-3}{*}{300 nm}          & \multicolumn{1}{c|}{$-1.59 \pm 0.33$}                & \multicolumn{1}{c|}{$-2.6 \pm 1.1$}                  & $-3.53 \pm 0.07$ \\ \hline
300 nm + Glycerol                 & \multicolumn{1}{c|}{$-1.19 \pm 0.19$}                & \multicolumn{1}{c|}{$-1.5 \pm 0.9$}                  & $-3.45 \pm 0.3$  \\ \hline
                                  & \multicolumn{1}{c|}{$-0.99 \pm 0.01$}                & \multicolumn{1}{c|}{$-1.53 \pm 0.13$}                & $-3.44 \pm 0.06$ \\
                                  & \multicolumn{1}{c|}{$-1.00 \pm 0.01$}                & \multicolumn{1}{c|}{$-2.30 \pm 0.12$}                & $-3.4 \pm 0.7$   \\
\multirow{-3}{*}{500 nm}          & \multicolumn{1}{c|}{$-0.99 \pm 0.01$}                & \multicolumn{1}{c|}{$-1.6 \pm 0.5$}                  & $-3.4 \pm 0.7$   \\ \hline
500 nm + Glycerol                 & \multicolumn{1}{c|}{$-1.53 \pm 0.09$}                & \multicolumn{1}{c|}{$-2.60 \pm 0.03$}                & $-3.51 \pm 0.02$ \\ \hline
                                  & \multicolumn{1}{c|}{$-0.2 \pm 1.4$}                  & \multicolumn{1}{c|}{$-2.75 \pm 0.16$}                & $-3.6 \pm 0.1$   \\
                                  & \multicolumn{1}{c|}{$-1.61 \pm 0.20$}                & \multicolumn{1}{c|}{$-2.63 \pm 0.08$}                & $-3.53 \pm 0.05$ \\
\multirow{-3}{*}{1300 nm}         & \multicolumn{1}{c|}{$-1.00 \pm 0.01$}                & \multicolumn{1}{c|}{$-1.6 \pm 0.5$}                  & $-3.4 \pm 0.7$  
\end{tabular}
\end{table}

\clearpage
\bibliography{Refs}

\end{document}